\documentclass[12pt,preprint]{aastex}

\shorttitle{A Deep Proper-Motion Survey in Kapteyn Selected Areas}
\shortauthors{Casetti-Dinescu, et al.}

\begin{document}

\title{A Deep Proper-Motion Survey in Kapteyn Selected Areas: I. Survey Description and 
First Results for Stars in the Tidal Tail of Sagittarius and in the Monoceros Ring}

\author{Dana I. Casetti-Dinescu\altaffilmark{1,2}, 
Steven R. Majewski\altaffilmark{3}, Terrence M.
Girard\altaffilmark{1}, Jeffrey L. Carlin\altaffilmark{3}, William F. van Altena\altaffilmark{1}, Richard J. Patterson\altaffilmark{3}
and David R. Law\altaffilmark{4}}

\altaffiltext{1}{Astronomy Department, Yale University, P.O. Box 208101,
New Haven, CT 06520-8101, (dana,girard,vanalten@astro.yale.edu)}
\altaffiltext{2}{Astronomical Institute of the Romanian Academy, Str.
Cutitul de Argint 5, RO-75212, Bucharest 28, Romania}
\altaffiltext{3}{Department of Astronomy, University of Virginia,  P.O Box 400325, Charlottesville, VA 22904-4325, (srm4n,jc4qn,rjp01@mail.astro.virginia.edu)}
\altaffiltext{4}{Department of Astronomy, Mail Code 105-24, California Institute of Technology, 
1200 East California Boulevard, Pasadena, CA 91125, (drlaw@astro.caltech.edu)}

\begin{abstract}

We describe a high-precision, deep (to $V \sim 19-21$) absolute proper-motion survey
that samples $\sim 50$ lines of sight in the Kapteyn Selected Areas along declination zones
$-15\arcdeg$,  $0\arcdeg$ and $15\arcdeg$.  In many fields the astrometric baseline reaches nearly
a century.  We demonstrate that these data provide typical per star precisions between $\sim1$ and 3 
mas yr$^{-1}$ to the above magnitude limits, with the absolute reference frame established by numerous
extragalactic sources in each survey field.  Combined with existing and ongoing
photometric and radial velocity surveys in these fields, these astrometric data will enable, among other things, 
accurate, detailed dynamical modeling of satellite interactions with our
Galaxy. In this contribution we describe the astrometric part of our survey and show
preliminary results along the trailing tail of the Sagittarius dwarf galaxy,
and in the Monoceros ring region.

\end{abstract}


\section{Introduction}

Modern, deep, uniform, large-area photometric surveys have shown unambiguously that the 
Milky Way outer halo contains accretion-derived substructure 
(e.g., Ibata et al. 2001, Newberg et al. 2002, Majewski et al. 2003, Rocha-Pinto et al. 2003, 2004,  
Vivas et al. 2004, Conn et al. 2005a). These 
structures of known or assumed remnants of satellite accretion,
have long-lived, coherent tidal features that can be used to 
model the Galactic gravitational potential (e.g., Ibata et al. 2001, Law et al. 2005), as well as the 
characteristics of the original satellite (e.g., Johnston et al. 1999).
However, such modeling studies have been limited by the meager
available kinematical data over large angles along the tidal features.
While radial-velocity programs have just recently begun to address this problem
for the few known Galactic tidal tails (Sagittarius - Majewski et al. 2004, Vivas et al. 2005, 
and the Monoceros ring - Crane et al. 2003), no systematic survey has begun 
to address the transverse (tangential)
velocities (i.e., absolute proper motions). 
Without this information, dynamical models remain 
poorly constrained, and therefore limited to describing merely a range of
possible events, rather than an accurate description of the real event.

The work described here is a proper-motion survey that provides
 high precision (1 to 3 mas/yr per well-measured 
star) absolute and relative proper motions down to a magnitude of 
$V \sim 19$, and for a few selected fields down to 
$V \sim 21$, in $\sim 50$ lines of sight 
in the Selected Areas (SA) designed by Kapteyn for Galactic structure 
studies in 1906. Current proper-motion surveys do not achieve this
precision at a similar magnitude limit and thus have limited capability to detect
and characterize distant halo substructure. To date there are only a few 
similarly deep, precise, pencil-beam-type proper-motion data 
sets that are primarily centered on globular clusters or dwarf spheroidals, 
and even fewer that are focused on Galactic field stars 
%such as traditional studies at the North and South Galactic poles 
(e.g., Chiu 1980, Majewski 1992, Guo et al. 1993, Dinescu et al. 2002).  
Obviously, ground-based studies of this
type are limited to parts of the sky where suitable first epoch astrometric data
exist.  Here we exploit a unique cache of nearly century old, deep photographic
plates having good scale, as well as other plate material collected in Kapteyn fields.

We intend to complement the new proper motion data with radial velocities, distances and 
metallicity estimates from our own photometric and spectroscopic work as well 
as from overlapping surveys such as 2MASS, SDSS, QUEST and RAVE. 

With this survey we aim to (1) determine the extent and orbital motion of the
highly obscured Monoceros ring-like structure above and below the 
Galactic plane, and explore its possible relation to other low latitude structures, 
such as the Canis Major (CMa) overdensity,
(2) characterize the transverse motion of the Sagittarius tidal streams,
(3) search for and characterize additional substructures in the halo of the
Milky Way and (4) determine the kinematical properties of numerous thick 
disk and halo stars in our fields as a function of Galactic position.

Future papers will include detailed kinematical analyses for specific
regions, while in this paper
we characterize the survey and show its potential in two 
specific regions where tidal streams have already been identified from other
sources. In the following Section we will describe the survey in detail. In Section 3
we show results in a few SA fields, and a brief summary is presented in Section 4.

\section{Survey Description}
\subsection{The Collection of Photographic Plates}
This survey is made possible by the visionary, now century-old Milky Way 
survey introduced by Kapteyn in 1906. Kapteyn devised the {\it Plan of 
Selected Areas} as a means to systematically study the Milky Way. 
The {\it Plan}, as 
originally envisioned, involved photometry, astrometry and spectroscopy 
of stars in 206 SAs collected at numerous observatories around the world
and focused on characterizing the ``sidereal world"; the perceived importance of this
grand effort in the early 20th century prompted the creation of 
International Astronomical Union (IAU) Commission 32: Selected Areas, 
as well as the Subcommittee on Selected Areas of IAU Commission 33: 
Structure and Dynamics of the Galactic System. Among the important early 
contributions to the SA program was the Carnegie Institution's systematic 
analysis of stellar photometry from Mount Wilson 60-inch plates of the 
139 northern accessible SAs by Sears, Kapteyn and van Rhijn (1930). 
The photographic plates used in this analysis were taken by Fath and 
Babcock with the 60-inch telescope between 1909 and 1912.
For 54 near-equatorial fields there exist deliberately matched 
(in area, approximate plate scale and depth) photographic plates taken with the 
Las Campanas Du Pont 2.5-m telescope by S. Majewski between 1996 and 1998.
This collection of photographic plates provides the opportunity for an
unprecedentedly deep, high-precision proper-motion survey that takes advantage of the excellent plate scale ($10.92\arcsec$/mm for the 2.5-m Du Pont, 
and $27.12\arcsec$/mm for the 60-inch Mt. Wilson) 
of the images taken in both epochs 
which span a $\sim 90$ year baseline. Each field of view is 
$40\arcmin\times40\arcmin$. Unfortunately, because the old 60-inch plates go only as deep as 
$V \sim 19$ for blue objects, there are very few galaxies to determine
the correction to absolute proper motion in a $40\arcmin\times40\arcmin$
field of view, and with those that are available being at
%. In addition these galaxies are at 
the plate limiting magnitude, and therefore
yielding rather poor centroids. Primarily for this reason, we have included in our 
proper-motion determinations the first
Palomar Observatory Sky Survey (POSS-I) 
plates as measured by both Space Telescope Science Institute (STScI, 
the Digitized Sky Survey - DSS) and by USNO. The POSS-I plates were 
taken in the early fifties
with the Oschin Schmidt telescope and have a plate scale of $67.2\arcsec$/mm.
While the scale and the digitization of the POSS-I plates (see Section 2.3)
are much poorer than those of the other two sets of plates, they offer
a 40-year baseline with the DuPont plates, and extend the proper 
motion limiting magnitude to $V \sim 20-21$.

Minimally, each field has an early epoch Mt. Wilson plate with two offset images, 
one from a 60 min exposure and one from a 3 min exposure. Many fields have a second early Mt. Wilson plate,
sometimes with two exposures, sometimes with one. For each field 
there are two recent-epoch  Du Pont plates taken in the 
blue (IIIa-J + GG385) and visual (IIIa-F + GG495) passbands.
These plates also contain a pair of offset exposures of about 60 min and 3 min integration.
The intermediate epoch 
POSS-I plates were taken in the blue (103a-O, no filter) and red 
(103a-E + RP2444) passbands, with typical exposure times of 10 and 50 minutes
respectively.

Finally, a handful of fields have KPNO Mayall 4-m prime focus photographic
plates (plate scale $18.6\arcsec$/mm) taken in the mid seventies by A. Sandage,
and mid nineties by S. Majewski.
The modern plates
were taken in the blue (IIIa-J+GG385) and visual (IIIa-F+GG495) passbands,
while the Sandage plates were taken in the blue (IIa-O+GG385/GG3), visual 
(IIa-D+GG495/GG11) and red (127-04+RG610) passbands. 

\subsection{Area Coverage}

In Figure 1 we show the location of the centers of the 
SA fields (open circles) on the sky in an Aitoff
projection. The filled circles represent the fields that also
have Mayall 4-m plates.
The survey samples three declination zones ($0\arcdeg$ and 
$\pm15\arcdeg$) and
the full range in right ascension, except for low ($|b| < 25\arcdeg$) 
Galactic latitude zones. The Galactic plane is represented with a grey
dash-dot line. There is one field, SA 29, which is at higher declination, 
and which has only 4-m plates. We are including this field because 
it falls within the QSO catalog determined from Data Release 3 (DR3)
of the Sloan Digital Sky Survey (SDSS) area (grey area in Fig.\ 1). 
%The SDSS area here is represented by the QSO catalog (Schneider et al. 2005)
%determined from the Data Release 3 (DR3). 
Although, for our astrometric 
reductions, we make use of the photometry from DR4, Figure 1 highlights
% used the more recent photometric DR4, in this plot we
%chose to show, besides 
the approximate SDSS footprint corresponding to the QSO catalog. 
Thus it can be seen that
for a good number of fields QSOs have already been identified that can be
used in setting the absolute proper-motion reference frame.

\begin{figure}[htb]
\includegraphics[scale=1.00,angle=-90,clip=true]{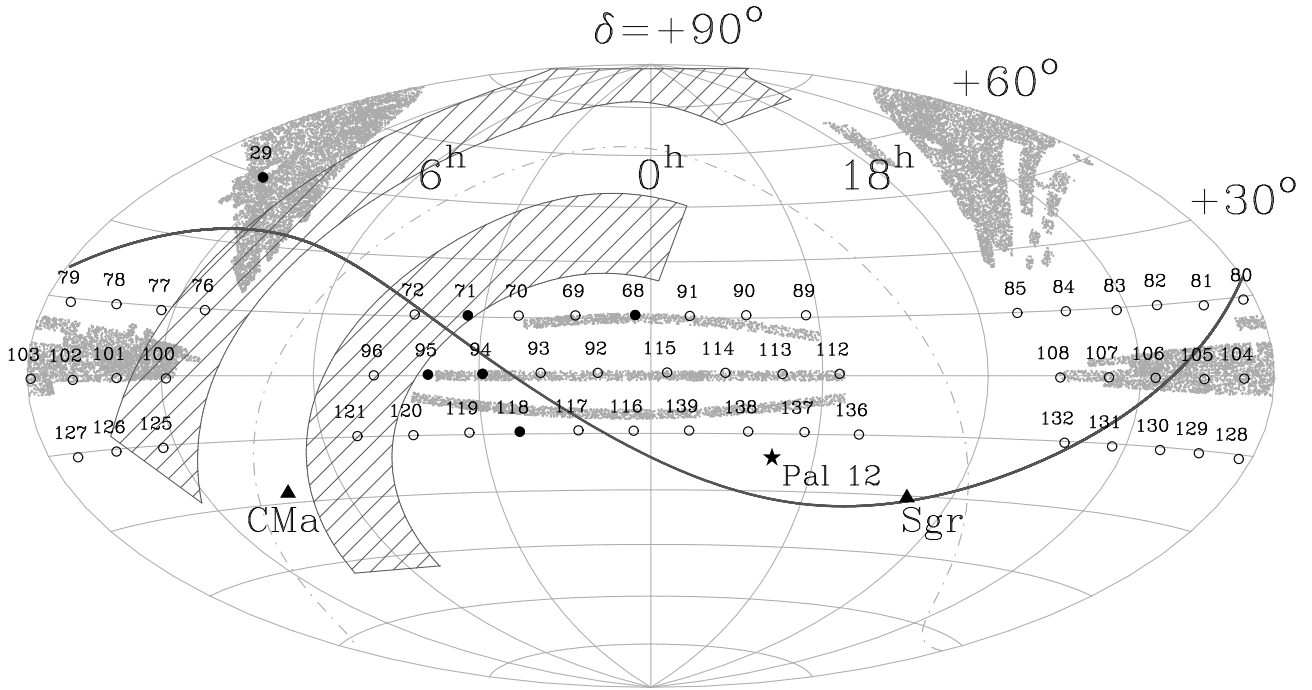}
\caption{Distribution of the SA field centers (open circles) in equatorial coordinates. Filled circles show the fields that have additional 4-m plates. The continuous line is the approximate orbital plane of Sgr, the dash-dot line is the Galactic plane. 
The grey area shows the SDSS coverage as given by the catalog of QSOs (Schneider et al. 2005). The cross-hatched bands represent the areas where the Monoceros
structure was mapped. The globular cluster Pal 12, Sagittarius's core and
 the Canis Major dwarf galaxy candidate are also marked.}
\end{figure}

We have indicated the location of the Sagittarius dwarf galaxy's (Sgr) center,
and that of globular cluster Pal 12, which is believed to have belonged to Sgr 
according to proper-motion data (Dinescu et al. 2000), 
surrounding field photometry (Mart\'{i}nez-Delgado et al. 2002) 
and chemical abundances (Cohen 2004). 
%SRM: You can also mark and mention Pal 2, which has been shown 
% (Lynden-Bell \& Lynden-Bell 1995, Majewski et al. 2004) to be a likely member 
% of the Sgr trailing stream.
The continuous dark line
shows Sgr's most recent orbit determination (Dinescu et al. 2005a). The orbit
is roughly indicative of Sgr's tidal streams that can be as wide 
as $10\arcdeg$, as Pal 12's location for instance suggests.
It is apparent that Sgr's southern, trailing arm, which is 
closer in distance to the Sun ($\sim 20-30$ kpc, Majewski et al. 2003) 
than much of the northern, leading arm, 
will be sampled in perhaps six SA fields (depening on the true stream width)
from RA = $0^h$ to $5^h$. In this paper 
we present results for SA 93 and 94. 

The second most well-known, putative halo substructure is the Monoceros stream or ring (``Mon"; 
Newberg et al. 2002, Ibata et al. 2003), a low Galactic-latitude structure 
which is represented here by the two cross-hatched 
bands above and below the Galactic plane. This area is drawn 
only approximately as based on
observations from Yanny et al. (2003), Ibata et al. (2003), Rocha-Pinto et al.
(2003),
Crane et al. (2003), Conn et al. (2005a), Martin et al. (2006).
The stellar overdensity discovered by Martin et al. (2004) in the constellation
of Canis Major and subsequently analyzed by other groups, may be a
distinct structure (but cf. Rocha-Pinto et al. 2006), and it has been suggested to be the core of 
the satellite that is responsible for the Monoceros stream (see, e.g. 
Pe\~{n}arrubia et al. 2005). We have marked the location of this 
structure (CMa) and we note that Conn et al. (2005b) claim, based on 
radial velocities, to have detected the Monoceros ring behind the 
CMa structure as well. At  first glance, between six and twelve SA 
fields are likely to sample the Monoceros ring. Fields SA 71 and 72 are
more problematical because both Sgr and Mon are expected in these areas 
(see Fig. 1). This paper presents results in SA 96, 100 and 101
that fall within/close to the Mon region.

Another equatorial survey that overlaps some of our proper-motion survey is the
QUEST survey (QUasar Equatorial Survey Team), which  
identifies RR Lyrae variable stars along the celestial equator (Vivas et al. 2004). Their 
most recent results (Duffau et al. 2006) indicate the discovery 
of a new halo structure in the constellation of Virgo that is not 
related to either Sgr or Mon (see also Juri\'{c} et al. 2006).
 Two of our SA fields, 103 and 104 are in the
area indicated by QUEST to sample the Virgo structure. 
Provided the depth of the plates allows it, we may be able to 
determine an absolute proper motion for this structure
in the near future. 

\subsection{Photographic Plate Measurements}
\subsubsection{Du Pont 2.5-m, Mt. Wilson 60-inch and Mayall 4-m Scans}
For each field, we start by fully digitizing one Du Pont, one 60-inch and one 4-m
plate. These initial coarse scans serve to build input lists for 
high-resolution scans.  
All of the scans, except for those of the Mayall 4-m plates
in SA 29, 71, 94 and 118,
were done with the Yale PDS Microdensitometer. 
The coarse scans of most of the 4-m plates were done with the University
of Virginia's microdensitometer. 
%The Du Pont and 60-inch plates coarse scans have a pixel size of $38\mu$m.
The size of the field is defined by the 10 inch $\times$ 10 inch Du Pont plate size, which corresponds to
$40\arcmin\times40\arcmin$.
For the 60-inch plate in each field, 
we digitize the same size area that matches the Du Pont field. This corresponds to
a 10-cm box located at the center of the Mt. Wilson plate. 
Stellar
images on the 60-inch plates are affected by coma, and outside this region 
they are practically unusable for astrometric purposes.
Based on these coarse scans, an input list of objects is determined
using the software package SExtractor (Bertin \& Arnouts 1996),
for each epoch. Long and short exposures on each plate are 
separated into two different lists. Then, each exposure on each plate is 
measured in a fine raster, object-by-object mode, with a pixel size of
$12.7\mu$m ($0.138\arcsec$) for the Du Pont plates, and $10\mu$m 
($0.275\arcsec$) for the 60-inch plates.
The input lists for the 4-m plates were made from the coarse 
scans done at UVa, and using the software package FOCAS 
(Valdes 1982)\footnote[1]{See also Valdes's 1993 FOCAS User's Guide, 
an NOAO document available at
ftp://iraf.noao.edu/iraf/docs/focas/focasguide.ps.Z}. These 4-m input catalogs
prepared earlier (Dinescu et al. 2002) were used to measure the 4-m plates
at Yale in a fine raster mode with a pixel size of $10\mu$m ($0.186\arcsec$).
The objects' positions, instrumental magnitudes, and other object parameters
were derived from the fine raster scans using the 
Yale 2D Gaussian centering routines (Lee \& van Altena 1983).
As is customary, a set of five to eight stars well-distributed over the plate
are repeatedly measured during the scan in order to monitor and correct
for thermal drifts during the scan.

From coordinate transformations of same-epoch, same-telescope
plates we obtain,
for well-measured stars, a centering precision of $1.2\mu$m (13 mas) per
single measurement, per star for the Du Pont plates. For the 60-inch
plates this number is $3.3\mu$m (90 mas), while for the 4-m plates,
it is $1.3\mu$m (24 mas).

\subsubsection{POSS-I Scans}
There are two readily available scans of the POSS-I plates:
those of the red plates done with a PDS machine at STScI,
widely known as the DSS, and those of both blue and red plates done at 
US Naval Observatory (USNO), Flagstaff Station
with the Precision Measuring Machine (PMM, see Monet et al. 2003 for
its description). 
The DSS scans are retrieved directly from the 
web\footnote[2]{http://archive.stsci.edu/cgi-bin/dss-form}
as the area in question is smaller than a degree on a side. 
The resolution of the DSS scans is $25\mu$m/pix ($1.7\arcsec$/pix).

Sections of the PMM scans were kindly made available to us by S. Levine at
USNO-Flagstaff. These scan sections are centered on the SA fields, and
cover $40\arcmin\times40\arcmin$. 
 The PMM is an 8-bit, fast measuring
machine that uses a CCD detector to take ``footprint'' images
of the photographic plate. The POSS-I plates were scanned at a resolution of
$13\mu$m/pix ($0.9\arcsec$/pixel). Each Schmidt plate is covered
by some 588 exposures with a field of view of $20\arcmin\times15\arcmin$ each.
The CCD footprints are assembled together in the
subsequent software by using an offset,
to provide the entire digitized sky (see details in Monet et al. 2003).
The ``stitching'' of the CCD footprints however is not perfect,
and thus position-dependent systematics are introduced
(see Section 2.5). Without access to the individual PMM footprints, a
method is required to correct the systematics in the assembled scans.
We have made use of the DSS scans, which were produced by a traditional
PDS measuring machine, to correct the PMM positions (see Section 2.5).
The SA field scans from DSS and USNO, for each POSS-I plate 
are processed as follows: 
Objects are detected with SExtractor and then re-centered with the 
Yale centering routines.
Coordinate transformations between overlapping plates for 
the DSS measurements indicate
a centering precision of $2.1\mu$m (141 mas) per single measurement, per star.
Similarly, for the USNO scans we obtain $\sim 2.6\mu$m (174 mas) for both 
the red and blue plates. Based on these values,  if we use only the
Du Pont and the POSS-I plates (a $\sim 40$-year baseline),
we can obtain a proper-motion uncertainty of $\sim 2$ mas/yr per
well-measured star.

\subsection{Photometry and Spectroscopy}

Here we will only briefly mention our campaign to obtain
photometry and spectroscopy in the SA fields; more details of
these observations will be presented in subsequent papers.

All SA fields have $UBV$ CCD photometry taken with
with the SITe\#1 2048$^{2}$ CCD on the Swope 1-m
at Las Campanas Observatory on the nights of UT 1997 December 23-31,
1998 June 19-26, and 1998 December 15-16.  The CCD field of view
covers $\sim 30\%$ ($22\arcmin\times22\arcmin$, $0.697\arcsec$/pixel) 
of the astrometric field.  Exposure times were 120,
200, and 900 seconds for the Johnson $V, B,$ and $U$ filters,
respectively, yielding data shallower by 1 to 1.5 magnitudes than the
typical Du Pont plates, for red stars, and by $\sim 0.3$ magnitudes for
blue stars.  Short exposures (5, 7, and 40-45 seconds in
$V, B,$ and $U$) were also taken to obtain photometry for the bright stars
in these fields.

For the astrometric reductions, we need $BV$ colors for all stars
to map out the color terms. These were obtained by calibrating
photographic instrumental magnitudes for each blue and visual 
Du Pont plate with CCD magnitudes.
The CCD magnitudes used in this process are not calibrated to 
the standard Johnson system because at the time the astrometric reductions
were done, the CCD photometry had not been calibrated yet. Nevertheless,
the CCD photometry helped to linearize the photographic magnitudes.

To date, the reduced CCD data cover only a 
handful of fields. Therefore, for the rest of the fields where 
there is coverage with SDSS, we have used the SDSS 
$(g-r)$ colors in the astrometric reductions.
In Figure 2 (left panel) we show the photographic, CCD-calibrated $B-V$ colors
versus the SDSS $(g-r)$ colors in SA 100. Dark symbols are stars, and
red symbols are galaxies.
The good color correlation justifies our use of the $(g-r)$ 
colors in the astrometry. The right panel shows the 
relationship between the SDSS $r$ magnitude and the photographic 
$V$ magnitude. This shows that the Du Pont plates reach a limiting magnitude of 
$r \sim 21$.

\begin{figure}[htb]
\includegraphics[scale=0.80,clip=true]{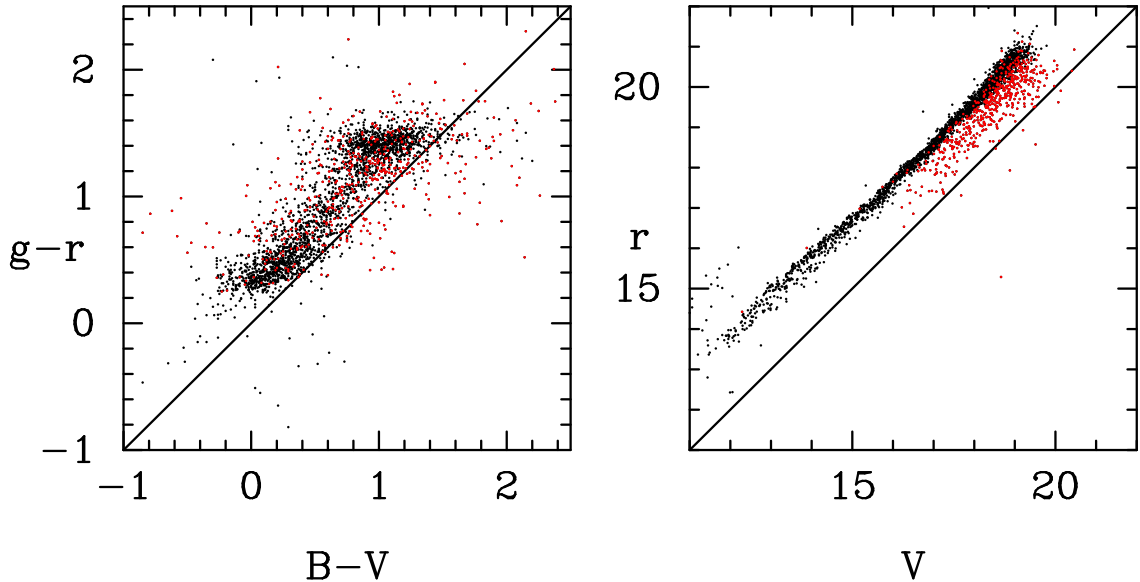}
\caption{SDSS $(g-r)$ colors as a function of photographic $(B-V)$
colors (left panel) and SDSS $r$ magnitudes as a function of photographic $V$
magnitudes (right panel). The photographic magnitudes are derived from the
scan of a Du Pont plate calibrated to CCD photometry. The $B,V$ magnitudes are
not calibrated to the standard Johnson system. Dark symbols represent the stars, 
while red symbols represent the galaxies.}
\end{figure}

For the fields where we have detected structure in the proper-motion
distribution as well as in the corresponding
SDSS color-magnitude diagram (CMD), we have started an observing
program with $Hydra$ on the 3.5-m $WIYN$ telescope 
in order to measure radial velocities.
So far, preliminary radial velocities have been obtained in SA 71 and SA 96.
These results will be presented elsewhere.

\subsection{Astrometry}
The Du Pont plates were precorrected for differential refraction 
(third-order refraction theory, Taff 1981) and for distortion
(see details in Dinescu et al. 2000). The distortion coefficients 
applied here were determined from the coordinate transformations of
eight visual and seven blue plates into the Second Naval Observatory
CCD Astrograph Catalog (UCAC2, Zacharias et al. 2004).
The distortion coefficient  for the visual plates is
$(-6.55 \pm 0.09) 10^{-8}$ mm$^{-2}$, and for the blue plates
is $(-7.02 \pm 0.06) 10^{-8}$ mm$^{-2}$. Between 100 and 200 stars in 
common with UCAC2 in the various SA fields were used to
model the coordinate transformations. The center of distortion 
was assumed to coincide with the tangent point. The tangent point
was determined by minimizing quadratic terms in the 
transformations of plate coordinates into the UCAC2 positions
(see e.g., Guo et al.  1993). The cubic distortion coefficients
 determined here, although
slightly smaller than that determined by Cudworth \& Rees (1991), are
in agreement within quoted uncertainties. The center of
distortion is determined with a precision of $\sim 0.5$ mm, or 5.5 arcsec.

The Mayall 4-m plates were precorrected for distortion using
the coefficients from Chiu (1976). The center of distortion is 
determined similarly to the process used for the Du Pont plates, i.e.,
by minimizing quadratic terms from a coordinate transformation into UCAC2.
This gives a distortion center known no better that $\sim 0.05$ mm (1 arcsec).
We have not precorrected the 4-m plates for differential refraction,
as this effect is much smaller than that of distortion; 
for the Du Pont plates these effects are similar in size (Dinescu et al. 2000).

The positions determined from the DSS scans of the POSS-I plates were used
to correct the positions determined from the PMM scans that are affected
by the assemblage of multiple CCD footprints (see Section 2.3). 
In Figure 3
we show the residuals from a coordinate transformation between 
PMM measurements and DSS measurements as a function of position.
The transformation includes up to fourth-order terms. The top two panels
show the residuals for the red plate (i.e., same red plate, two
different measurements, PMM into DSS) and the bottom two for 
the blue plate (i.e., blue PMM measurements
into red DSS measurements) as labeled.
The right panels show the residuals
after the correction was applied. The correction for each object is derived 
by taking a local average (of some 20 neighbors) from a 2-d  map of the residuals
shown in Fig. 3.

\begin{figure}[htb]
\includegraphics[scale=0.90,clip=true]{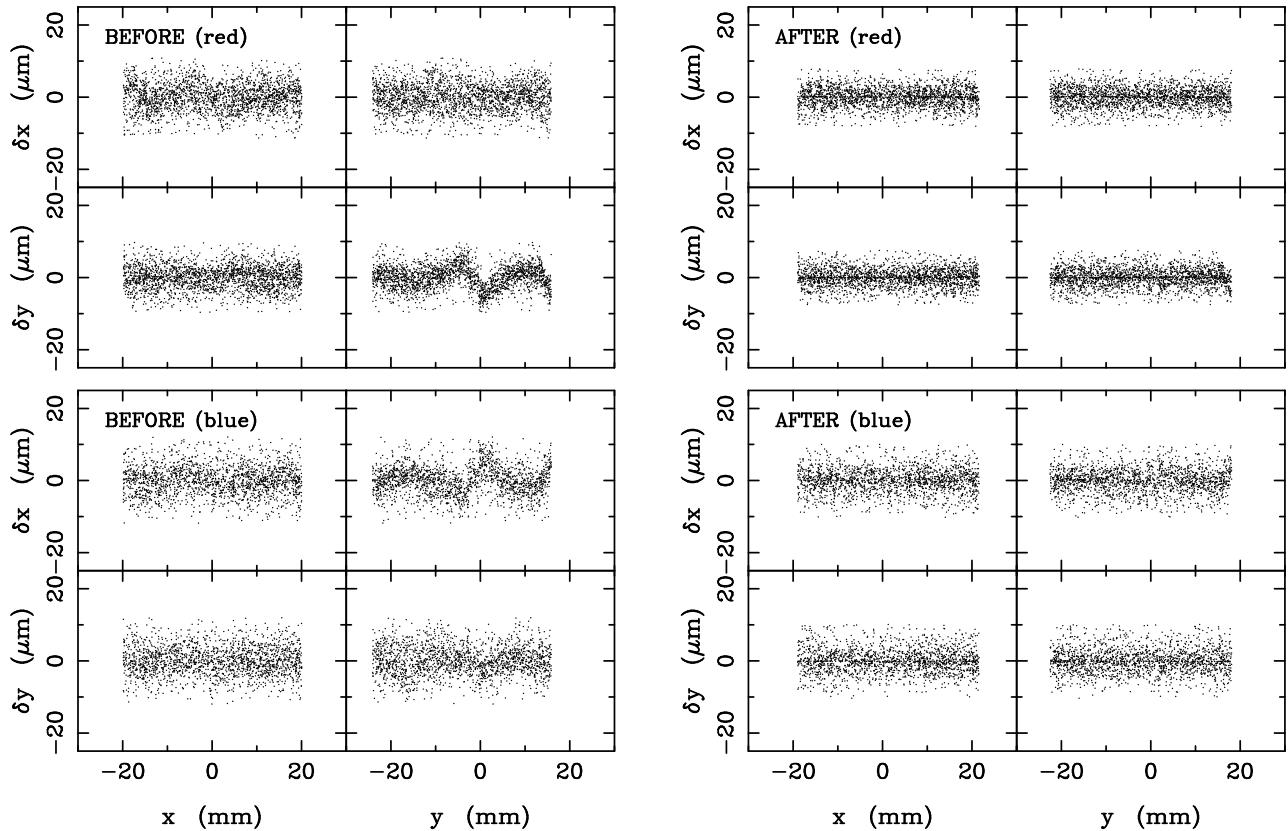}
\caption{Residuals from a transformation of the PMM measurements into the DSS 
measurements. The first two rows are for PMM measurements of a red plate
transformed into DSS measurements of the same red plate. The bottom two rows
are for PMM measurements of a blue plate transformed into DSS measurements
of a red plate. Left panels show the residuals before a 2-d correction 
was applied, and right 
panels after the correction was applied (see text).}
\end{figure}

It is well known that astrometry from the wide-field POSS-I plates, 
which were taken with a Schmidt telescope, is strongly affected 
by position and magnitude-dependent systematics (e.g., Morrison et al. 
2001 and references therein). These are particularly large and 
difficult to model when an entire plate of $6.5\arcdeg\times6.5\arcdeg$ 
is being mapped into astrometric catalogs. For our purposes however,
we are using only small regions on a plate, which can be mapped into one another
and into an external system, such as the Du Pont plates for example,
by using up to third order polynomials. As for magnitude dependent
systematics (i.e., the magnitude equation) , 
these are minimized by using stars within a relatively
narrow magnitude range,
toward the faint end of the plate limit: $V\sim 15$ to 19.

The old 60-inch plates were not precorrected in any manner
as they are the least well-understood.
We are aware that coma is present from the shape of
the images on the plates, and that it affects the astrometry
from coordinate transformations
of long into short exposure. In the following Section we describe
the way in which these plates were incorporated into the
proper-motion determinations.
 
\subsection{Proper-Motion Determinations} 
The lists of coordinates from the 60-inch, POSS-I and 4-m
plates are matched with the list from the Du Pont plates.
The matching radius is 2 arcsec. This basically limits our catalog to objects 
with proper-motions smaller than $\sim 50$ mas/yr.
Although stars with proper motions larger than this value can be recovered
from the current lists of positions for all measured objects on each plate, 
it is not our immediate goal to do so.

The proper-motions are determined differentially, by adopting 
one plate as the master plate, into which all the others are mapped.
The master plate is chosen from the Du Pont plates, which are
best understood and modeled. During long exposures, guiding-induced
magnitude-dependent systematics (also known as the magnitude equation, 
see e.g., Majewski 1992, Guo et al 1993, Girard et al. 1998) 
affect practically all photographic material. 
However, the Du Pont plates have a short exposure, and we have 
assumed that this exposure is not strongly affected by the magnitude equation.
Therefore we have transformed the long exposures into one short exposure,
and the magnitude-dependent trend of the residuals was used to 
correct the long-exposure measurements. This essentially corrects
the bright stars ($V \le 16$), over which the trend is apparent.
One of the magnitude-equation corrected Du Pont long-exposures is then chosen
as the master plate. The remaining plates are then transformed into 
the master plate using up to fourth order
polynomial coordinate transformations and linear color terms. 
These high-order geometric terms are present most likely due to our 
inability to accurately determine the center of distortion (see Section 2.5). 

Preliminary proper motions are calculated based on the Du Pont and 
POSS-I plate measures, and, for a handful of fields, from Du Pont 
and 4-m plates. The transformation of POSS-I plates into the Du Pont system 
occasionally requires third-order polynomials in the coordinates, 
and linear color terms. 
The derived preliminary proper motions allow propagation 
back in time to the epoch of the 60-inch plates, and
thus the 60-inch plates are tied to the system of the Du Pont 
master plate. These transformations include third order polynomials
in the coordinates,
coma terms and linear color terms.
New proper motions are then calculated from the entire 
set of plates and one more iteration is performed to obtain the final
values. The reference system that is used to determine the
plate transformations consists of faint stars, varying in number
from a couple hundred to a couple thousand, depending upon the 
%plates' depth and the 
Galactic latitude of the field and the depth of the corresponding plates. The resulting
proper motions are relative; the correction from relative to absolute 
proper motions is derived from the offset defined by the mean proper motion of QSOs and galaxies
in the field.
We note that the Du Pont and 60-inch short exposures
provide measurements of stars
as bright as some of the faintest $Hipparcos$ stars ($V\sim 8$).
However, we will not rely on any of the stars at the bright end
to determine the correction to absolute proper motions, because
we believe our proper-motions at the bright end ($V < 15$) are
affected by unaccounted for magnitude-dependent biases. 
The proper motions of galaxies and the QSOs are also used at each
iteration to check magnitude and color-dependent systematics. 
In some of the areas that have 4-m plates, galaxies were used to
correct small trends with color and magnitude left in the 
first-iteration set of proper motions.
QSOs and galaxies are selected from the SDSS (Schneider et al. (2005) and 
DR4 classification). According to the SDSS documentation, the
galaxy classification is quite reliable down to $r = 21$.
For areas that do not overlap with SDSS we have selected galaxies from
a visual inspection of the deepest Du Pont plate.

We calculate a proper motion for each object that has at least three measurements 
separated in time by at least $\sim 40$ years ($\sim 20$ years for 
fields that include 4-m plates).  The proper motion is calculated for
each object from a linear least-squares fit of positions as a function of
plate epoch. The formal proper-motion uncertainty is given by the
scatter about this best-fit line. Measurements that differ
by more than $0.2\arcsec$ from the best-fit line are excluded.
Objects that  have only three measurements should however be considered
with caution because they may have unrealistically small formal uncertainties.

\subsection{Proper-Motion Uncertainties}
There are  two ways to estimate externally the proper-motion
uncertainties: 
(1) by direct comparison with another high-quality
proper-motion catalog and (2) from the proper-motion scatter of 
objects that have no or negligible intrinsic proper-motion dispersion.
For the first test, we compare our proper motions with those in the Munn et al. (2004)
 catalog in five SA fields. We remind the reader that the Munn et al. (2004)
proper-motion catalog was made by combining USNO-B (Monet et al. 2003) 
with the SDSS (DR1). Munn et al. (2004) have used SDSS galaxies to correct 
for position-dependent proper-motion
systematics and to place the proper motions on an absolute reference frame.
In Figure 4 we show proper-motion differences (i.e.,
 our relative proper motions minus proper motions from Munn et al. 2004)
as a function of magnitude. Units for proper motions are mas/yr
throughout the paper. The left 
panels show proper-motion differences along right ascension (RA), and the 
right panel along declination (Dec). It is apparent that, in all fields, there
is a larger scatter in the RA proper-motion differences, than in 
Dec. Proper-motion differences plotted as a function of
positions and colors show that the larger scatter in RA is not due to
systematics related to these quantities. 
Our proper motions for galaxies and QSOs 
(i.e., objects with no intrinsic proper-motion dispersion) 
do not indicate that our measurements are consistently poorer in the RA
direction than in Dec (see below). Interestingly,
the histogram  of QSO proper motions in Munn et al. (2004, their Figs 1 and 3)
shows that the RA proper-motion dispersion is larger than that in
Dec, in agreement with our findings. 
From coordinate transformations of the modern Du Pont plates 
directly into SDSS positions in some thirteen SAs, 
we do not find indications that
the positional error in RA is larger than that in declination.
This leads one to conclude that the positional precision 
in the USNO-B catalog is poorer in RA than in declination.
Indeed, Figure 1 in Munn et al. (2004), which shows the distribution of 
QSOs' proper motions in RA and Dec for both 
the USNO-B catalog and the new Munn et al. (2004) catalog,
indicates that the scatter in RA is larger than that in Dec
in both catalogs.

\begin{figure}
\includegraphics[scale=0.90,clip=true]{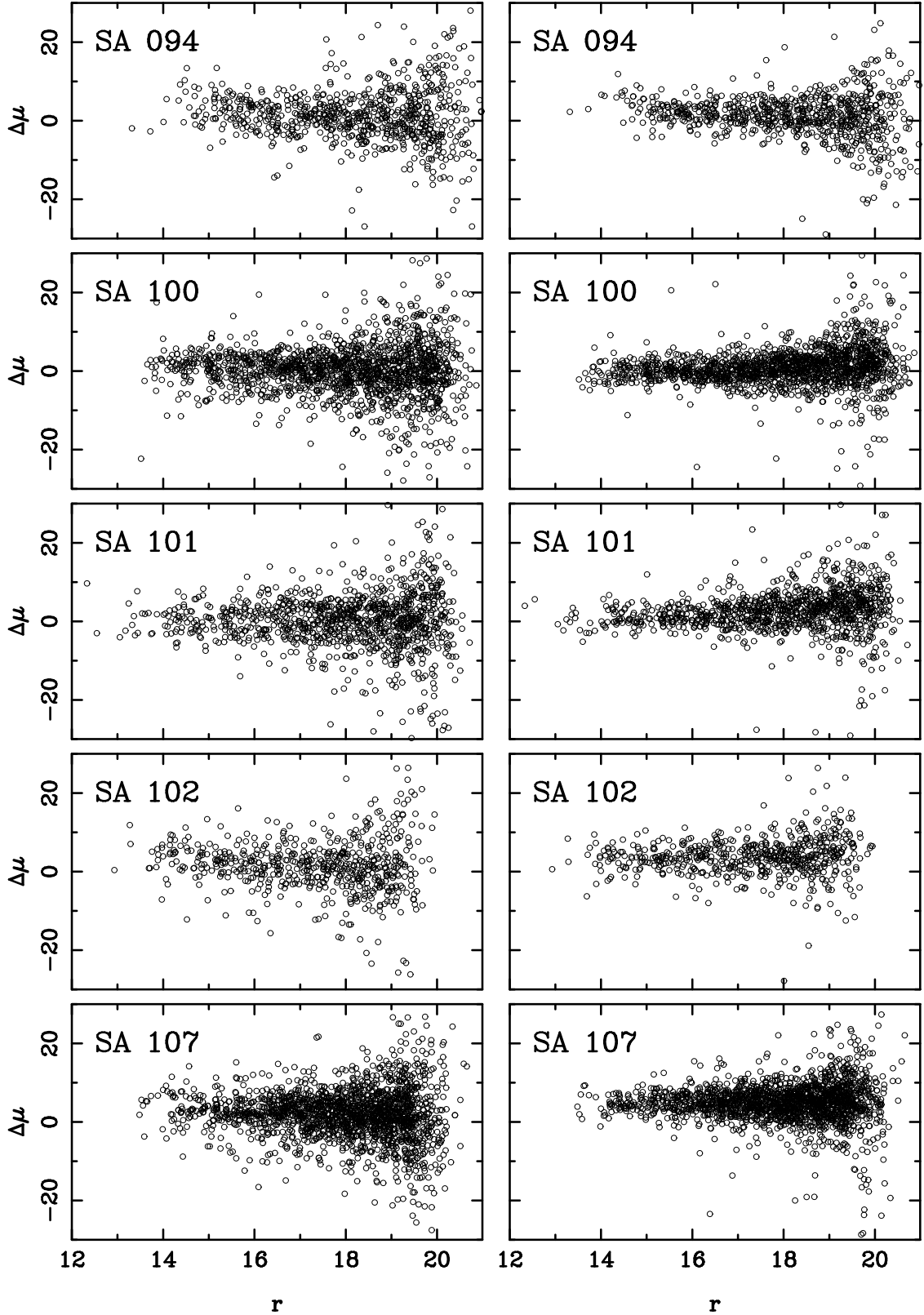}
\caption{Proper-motion differences (ours minus those in the Munn et al. 2004 catalog) as 
a function of magnitude in five SAs. The left panels show
the RA direction, right panels the Dec direction.}
\end{figure}

From our comparison with the Munn et al. (2004) catalog, 
we obtain a scatter of $\sim 3.8$ mas/yr in RA and $\sim 2.8$ mas/yr
in Dec , for $r < 18$.
Munn et al. (2004) quote uncertainties of
3.6 and 2.8 mas/yr in RA and Dec respectively for the 
same magnitude range.
This indicates that our uncertainties are substantially smaller.
The fact that we see in the 
proper-motion differences (Fig.\ 4) the larger scatter in RA than 
in Dec which is characteristic of the Munn et al. (2004) catalog, 
implies that the errors in the latter catalog dominate the scatter in
the proper-motion differences. 
  
Our second test uses galaxies, QSOs, 
stars in known streams or in clusters, i.e., objects for which the
proper-motion dispersion reflects only the measurement uncertainty.
In practice, galaxies have larger proper-motion uncertainties than 
stars due to poor centering of their fuzzy, low-gradient image profiles, and 
there are very few QSOs in each SA field. Thus these
uncertainty estimates should also be viewed as conservative numbers.
In Figure 5 we show relative proper motions of galaxies (open circles) 
and QSOs (filled red circles) as a function of magnitude
for the same SA fields shown in Fig. 4 (left panel is for RA, right panel for
Dec). Clearly SA 94 is better 
measured, and this is because it includes 4-m plates.

\begin{figure}
\includegraphics[scale=0.90,clip=true]{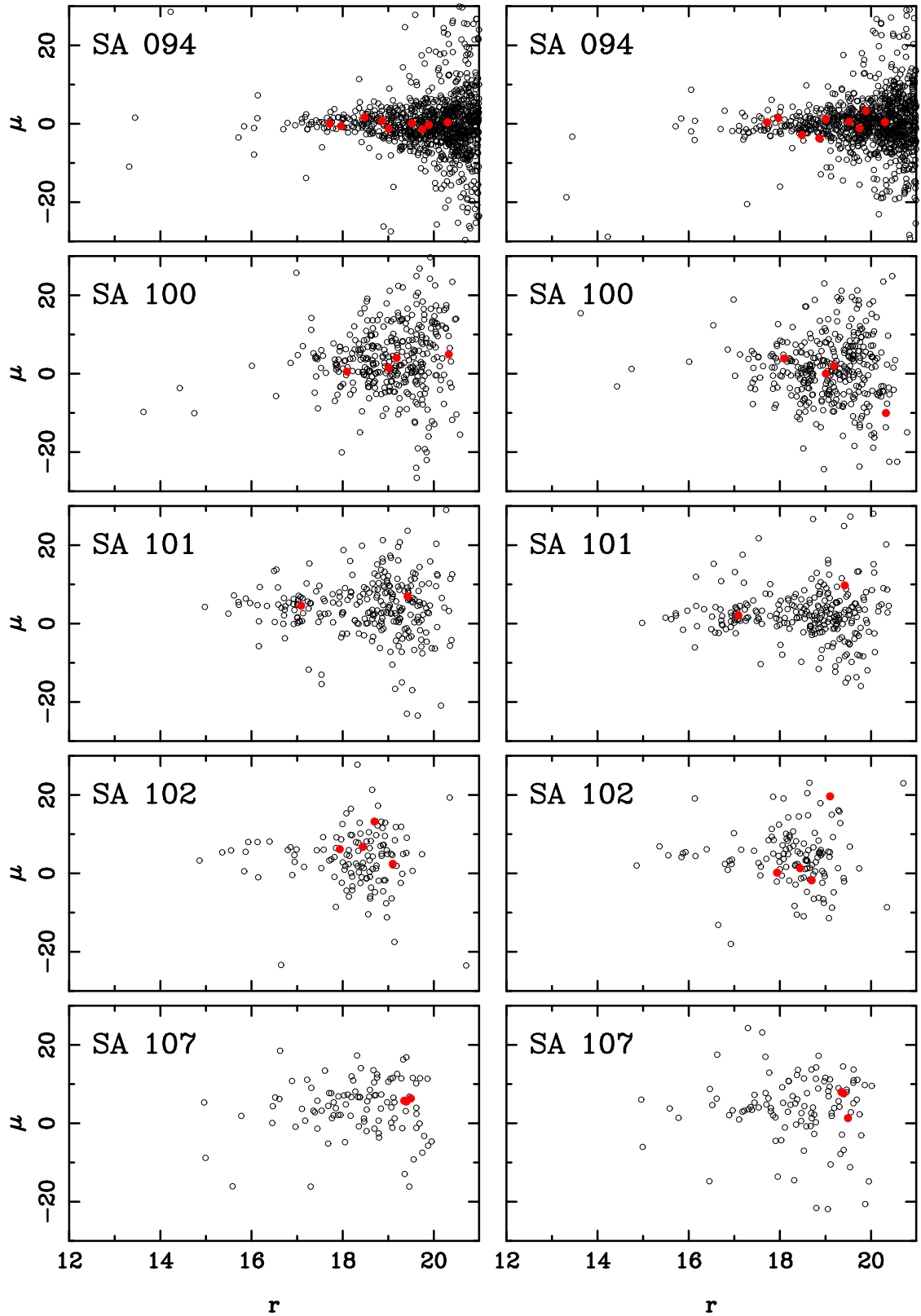}
\caption{Relative proper motions of galaxies (open circles) and QSOs
(filled red circles) as a function of magnitude for five SAs. Left panels show
the RA direction, right panels the Dec direction.}
\end{figure}

Figure 5 indicates that the proper-motion uncertainty varies considerably
with magnitude, and there may be slight 
variations from field to field. These variations 
are primarily due to the image quality of the old POSS I plates.
For a typical field, we obtain between 3 and 5 mas/yr for
reasonably well-measured galaxies, while for fields that
have 4-m plates, we obtain 2 to 3 mas/yr. Most fields have of the order of
100 galaxies, therefore the uncertainty in the correction to absolute
proper motion is $\le 0.5$ mas/yr. For the QSOs, it is more difficult
to reliably estimate uncertainties, due to small number statistics and 
the fact that most QSOs are toward the faint limit of the survey.
However, for SA 94, we obtain between 1 and 2 mas/yr for well-measured 
($r \le 19$) images.
Proper-motion uncertainty estimates as derived from 
``known'' tidal structures will be given in the following Section, where
we present the tidal tail results.

\begin{figure}[htb]
\includegraphics[scale=0.60,angle=-90,clip=true]{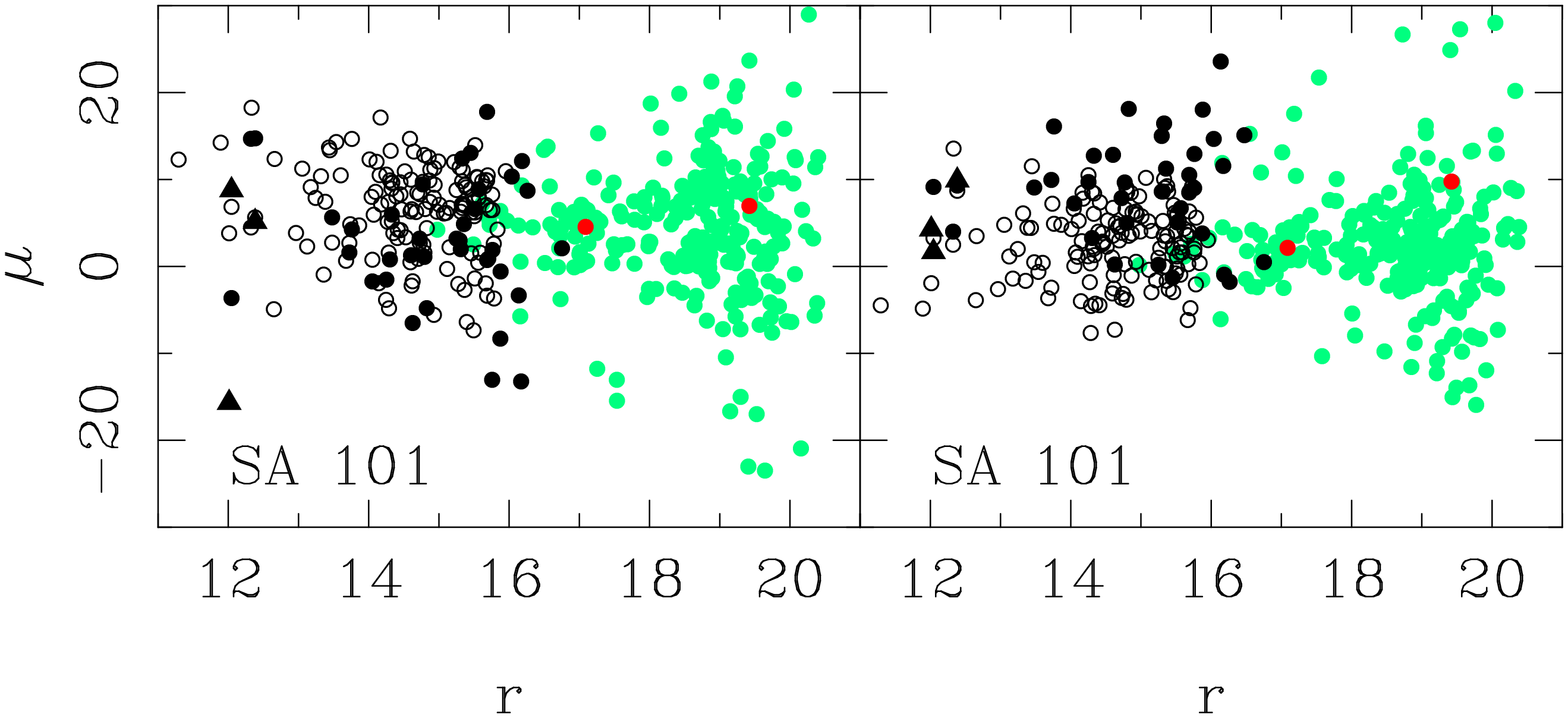}
\caption{Correction to absolute proper motion as a function of magnitude
as given by QSOs (red circles), galaxies (green circles), NPM1 stars (black filled circles), 
UCAC2 stars (open circles) and Tycho2 stars (filled triangles).
As in the previous two figures,  the left panel shows
the RA direction, the right panel shows the Dec direction.}
\end{figure}

Finally, in Figure 6 we show the correction to absolute proper 
motion as given by a number of calibrating objects for SA 101.
Most of the fields show similar characteristics.
Toward the faint end, the galaxies (green symbols) and QSOs (red
symbols) dominate. At the bright end we show the correction to absolute proper
motion as given by proper-motion differences of
Tycho2 (Hog et al. 2000) stars (filled triangles), 
UCAC2 stars (open
circles), and NPM1 (Lick Northern Proper-Motion Program, Klemola et al. 1987) 
stars (filled circles).  For all of these three catalogs,
the stars matched with our survey are at the faint limit of those catalogs,
so that these are stars with the largest measurement errors in these other
catalogs. Indeed, the proper-motion scatter is between 4 and 6 mas/yr,
poorer than that of our measured galaxies. This is one reason
why we have not used Tycho2, UCAC2 and NPM1 stars to calibrate to
absolute proper motions. The second reason is due to obvious offsets
between the zero point as determined by galaxies and QSOs in our survey,
and that determined based on brighter stars in these external catalogs. This 
is due to residual magnitude-dependent systematics in either or
both the listed catalogs and 
in our survey. Since the magnitude range of galaxies and QSOs better matches
that of our survey stars, using these to establish the abolute proper motion
correction minimizes 
magnitude-dependent systematics.
We will use both galaxies and QSOs in the determination of the 
absolute proper-motion correction.

One more question remains: Why not use stars in the Munn et al. (2004)
catalog as astrometric zero-point calibrators? From comparisons of the zero point as determined
by galaxies and by the Munn et al. (2004) stars, we find 
differences between 0.2 and 5 mas/yr. The scatter in the difference is 
2 mas/yr.
The absolute proper-motion calibration in Munn et al. (2004) is also
based on SDSS-selected galaxies, and their first epoch material is in our
time series.
Nevertheless, differences appear in the
two independently determined catalogs. We feel our proper motions
are more reliable for two reasons: (1) the modern epoch consists of precise
measurements of the high quality Du Pont plates, and (2) 
the old epoch, i.e., the POSS-I scans were re-reduced here in a very careful
manner to  minimize position-dependent systematics (see Section 2.5),
while the Munn et al. (2004) proper motions necessarily had to
rely on the batch-reduced PMM scans of the POSS-I plates for
nearly $10^{9}$ objects.
For these reasons, as well as because of the effect seen in Figure 4, we will use
our internal zero-point calibration based on galaxies and QSOs.

\subsection{Completeness}

\begin{figure}
\includegraphics[scale=0.60,clip=true]{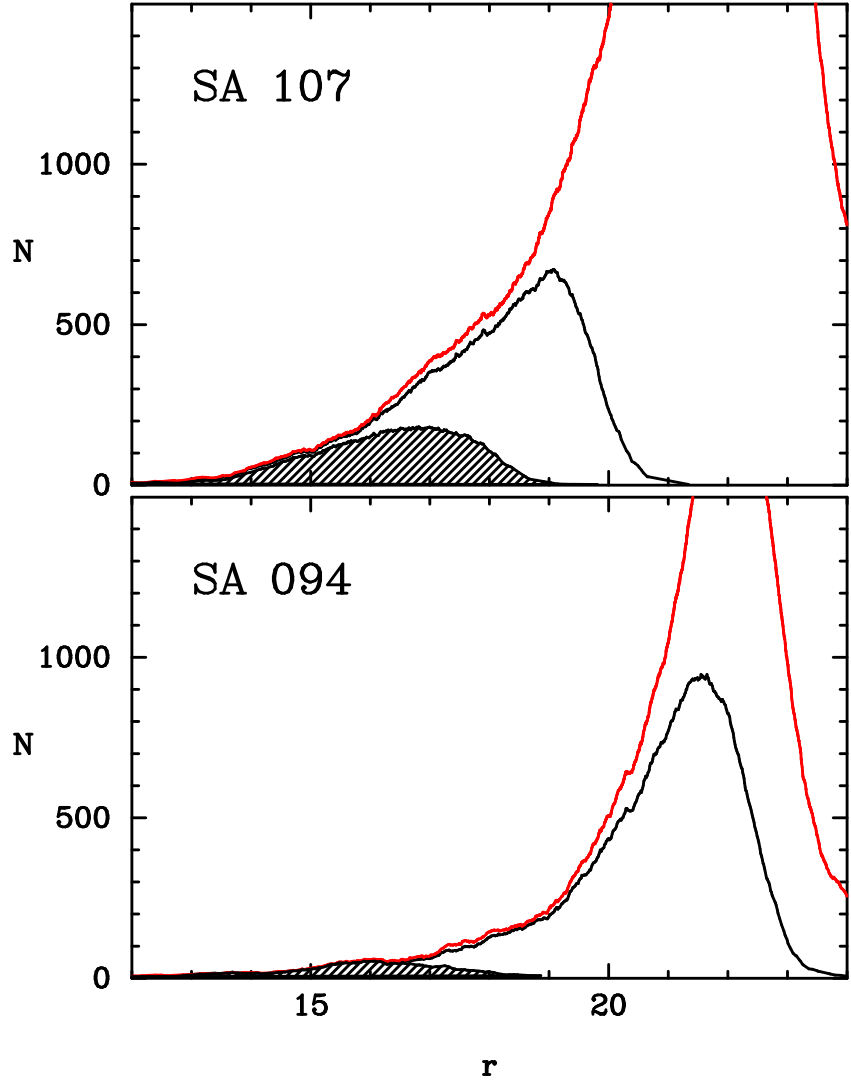}
\caption{Distribution of objects as a function of magnitude 
for SA 107 and SA 94. The SDSS catalog is represented with a red line, 
our survey  with a black line. The hatched area is  
our survey, only  for objects measured on the old 60-inch 
plates. }
\end{figure}

In Figure 7 we show the distribution of objects detected
as a function of magnitude
in two fields, SA 107 a typical field in our survey, and SA 94, a
deep field that includes 4-m plates. The SDSS distribution is represented
with a red line, our survey with a black line, and the hatched area is
that fraction of our
stars that were measured on the old 60-inch plates. 
For a typical survey field, our catalog is nearly complete to $ r = 19 $, and 
this magnitude limit varies slightly from field to field depending 
on the deepness of the POSS-I plates. Figure 7 shows that the SDSS 
distribution and our catalog distribution follow each other closely.
A difference between the SDSS and our catalog counts is however 
present between $r = 17$ and 19.  This difference indicates that
SDSS has 6 to $9\%$ more counts than our catalog.
For these two fields, we have checked by hand
the objects that appear in SDSS and do not appear in our catalog.
We found that the absence of these objects from our catalog is  
due to two reasons: (1) from its construction (Section 2.6)
our catalog misses high (greater than $\sim 50$ mas/yr) 
proper-motion stars, and (2) SDSS has a non-negligible
number of spurious detections near bright stars that, upon inspecting the  
SDSS images, appear to be located on diffraction spikes. 
We note that the latter objects have 
SDSS flags that qualify them as real, primary detections
(i.e., flag ``GOOD'').
To correctly quantify the contribution of each of 
these effects is not a trivial matter. 
If we assume however that the SDSS represents 
the true counts, we then obtain a conservative estimate of our 
completeness limit which is greater than $90\%$ to $ r= 19$. The limiting
magnitude at $50\%$ completeness is $r \sim 19.5 $. For fields that include 4-m plates, the
catalog is near-complete (greater than $90\%$) 
to $r \sim 21 $ and reaches a limiting magnitude 
of $ r= 22 $ at $50\%$ completeness.

\section{Results}

In what follows we show proper-motion results in SA fields containing
some currently known halo substructures. Quantitative analyses that involve
absolute proper motions combined with other information and
with models will be presented elsewhere.

\subsection{Sagittarius Tidal Tails}

Based on Hess diagrams, Newberg et al. (2002) have identified 
overdensities of F-colored stars related to debris from the
Sagittarius dwarf in the regions labeled S167-54-21.5 (their Fig.\ 7),
and S341+57-22.5 (their Fig.\ 5). 
The latter region includes our SA 105 and 106.
We find however very little evidence of clumpiness in the proper-motion diagram
because Sgr's main sequence turnoff is at very faint magnitudes ($r = 22.3$).
A small number of Sgr red clump stars at $r \sim 19.5$ may be present in 
SA 105 and 106, but these are at the faint end of our survey and
will have large proper-motion uncertainties. 
This SDSS region samples distant parts of 
the leading tidal tail of Sgr (see Fig. 1 and e.g., Majewski et al. 2003).
Thus we no longer consider these fields here.

The SDSS region S167-54-21.5 includes 
SA 93 at $(l,b) = (154.2\arcdeg,-58.1\arcdeg)$
and SA 94 at $(l,b) = (175.3\arcdeg,-49.2\arcdeg)$, and it samples 
the trailing tidal tail of Sgr, which is generally closer to the Sun than the
leading trail (e.g., Majewski et al. 2003).
Yanny et al. (2003, 2004), Majewski et al. (2004) and Law et al. (2005) 
 confirm with radial velocities 
the presence of Sgr debris in this region.
In Figure 8 we show the SDSS CMDs and relative proper-motion
distributions for SA 94, 93 and 107. The middle and
right hand panels show proper motions: the $x$ direction corresponds to 
proper motions along RA, and the $y$ direction to Dec.
Each row represents one
field.  SA 107 located at $(l,b) = (5.7\arcdeg,+41.3\arcdeg)$  
was chosen for comparison with the other two fields, and
it is roughly symmetrically placed in the Galaxy with SA 94. 
The reddening is relatively low in all fields: for SA 93 it is $E(B-V) \sim 0.03$,
for SA 94 it is $0.09$, and for SA 107 it is 0.11 (Schlegel, Finkbeiner, \& Davis 1998).
The CMDs show objects in the SDSS that were matched with our proper-motion
survey. Galaxies according to the SDSS classification were eliminated.
The magnitudes and colors are not dereddened. For SA 94, where we have better statistics
for the QSOs than in the other fields, we  highlight the
QSOs with red symbols so that their proper-motion distribution can be compared to
that of stars. 

The middle panels show the proper-motion distribution of blue stars
($ 0.2  < g-r  < 0.8$) chosen to represent the
turnoff of thick disk and halo stars, and the right panels that of red stars
($1.2 < g-r < 1.7$) chosen to represent more nearby, disk dwarf stars.
Sgr's turnoff is visible in the CMDs of SA 94 and 93, at $r \sim 21 $.
SA 107 is also located in the 
SDSS region S6+41-20, which Newberg et al. (2002)
find is most consistent with a smooth halo/thick disk population.
Because of its direction toward the inner Galaxy
 and the fact that the 
scale length of the thick disk is 3-4 kpc (e.g., Juri\'{c} et al. 2006),
SA 107 samples more thick
disk stars than SA 94. The proper-motion distributions
show the intrinsic dispersion convolved with the proper-motion
measurement uncertainty, which increases with magnitude.

\begin{figure}
\includegraphics[scale=0.95,clip=true]{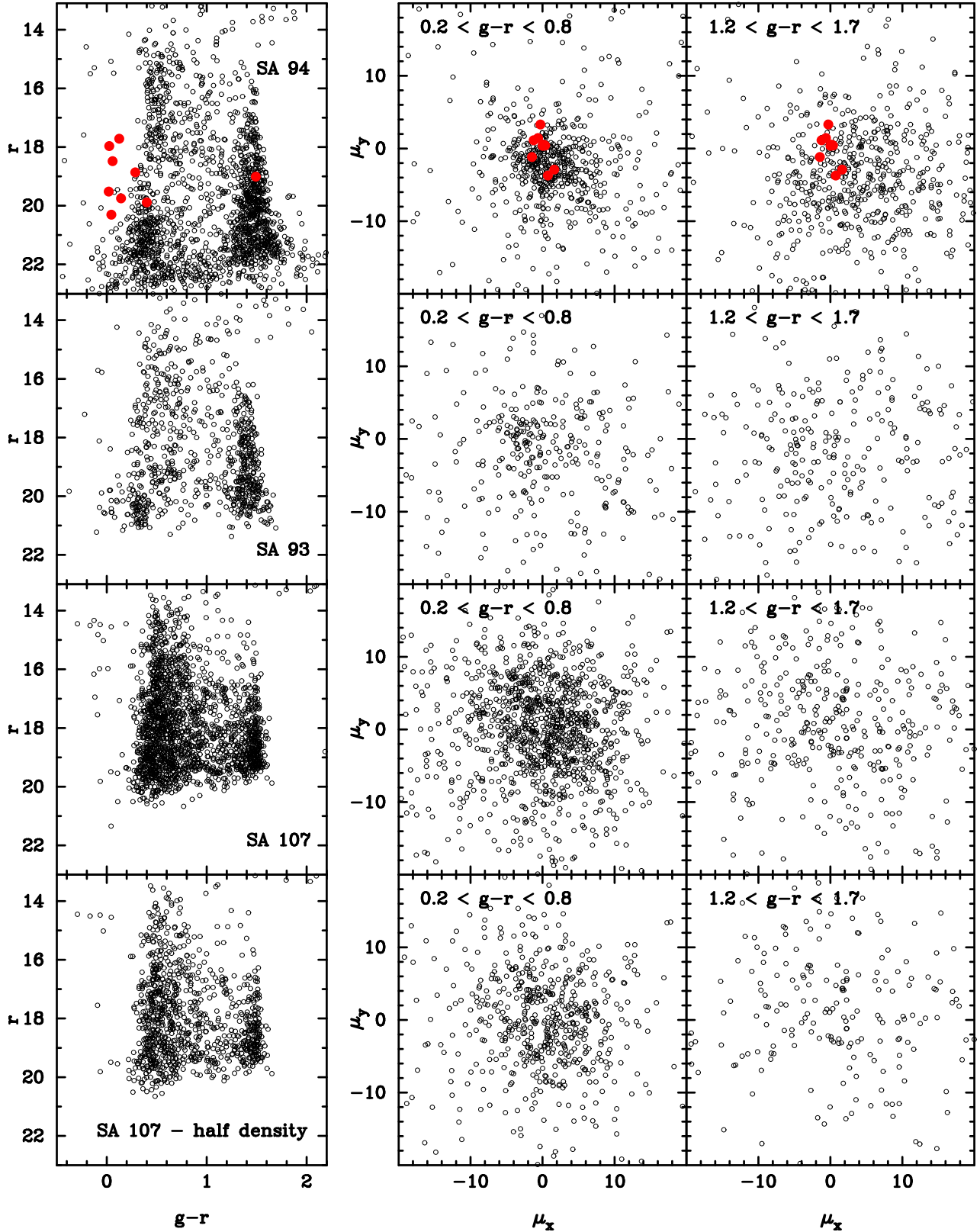}
\caption{CMDs (left panels) and relative proper-motion diagrams 
(middle and right panels) for stars and QSOs in
SA 94, 93 and 107. The middle and right panels show the proper motions for blue
$(0.2 < (g-r) < 0.8)$ and red $(1.2 < (g-r) < 1.7)$ stars respectively.
In the top row, QSOs are highlighted with red symbols for comparison
with stars' distributions.
The bottom row shows SA 107, where only every
other star was plotted to mimic a lower density field.}  
\end{figure}

By comparing the Figure 8 proper-motion distribution of presumably nearby red stars 
in all fields, it is
clear that SA 94 is a much better measured field --- a 
result of having 4-m plates in its time series data.
The proper-motion uncertainties for SA 93 and SA 107 are however comparable
to one another.
The proper-motion distribution in SA 107 shows a clear kinematical distinction
between blue stars --- i.e., distant thick disk and halo stars near the main sequence turnoff ---
and red, nearby dwarf stars.
%. Indeed the larger proper-motion dispersion of red stars compared to that of blue stars
%is representative of the kinematical distinction of the two populations.
The fourth row shows the same diagrams as in the third row, but with only half the stars
in SA 107 plotted  so as to approximately match the number of
stars in SA 94 in the appropriate magnitude range.
It is evident  that the blue stars in SA 94 and 93 show more concentrated clumping
in proper-motions due to the presence of a population with a dispersion much
tighter than that expected for random thick-disk/halo stars (i.e., SA 107).

This identified proper-motion clumpiness together with a sudden overdensity of stars
at a particular magnitude in the CMD is a clear signature of distinct substructure in the halo,
presumably dominated by Sgr tidal debris.
In SA 94, the best measured field,
the proper-motion scatter  of the clumped, blue stars is
$\sim 1.8 - 2.0$ mas/yr.
This number is determined primarily by faint stars ($r = 21$) and
reflects our proper-motion uncertainty. 
In this contribution we present the mean absolute proper motion of candidate
Sgr debris in SA 94 and 93.

In SA 94, the correction to absolute proper motion is given by the error-weighted
mean of two determinations: that with respect to QSOs and that with
respect to galaxies. 
The former determination is the average of nine QSOs and it
is $\mu_{\alpha}cos\delta = -0.08 \pm 0.32$ mas/yr and $\mu_{\delta} = -0.07 \pm 0.73$ mas/yr. 
The latter includes 885 galaxies with magnitudes between 
$r= 15$ and 21 that
have measurements on at least four plates and proper-motion values less than 20 mas/yr
in both coordinates. Their average proper motion
is $\mu_{\alpha}cos\delta = -0.39 \pm 0.18$ mas/yr and $\mu_{\delta} = -0.01 \pm 0.19$ mas/yr. 
The error-weighted average of these two determinations is: $\mu_{\alpha}cos\delta = -0.32 \pm 0.16$ mas/yr 
and $\mu_{\delta} = 0.00 \pm 0.18$ mas/yr.
Candidate Sgr stream stars were selected based on the SDSS CMD (Fig.\ 8)
to belong to the turnoff region. Only stars with measurements on at least four plates and
with $r$ magnitudes brighter than 21.7 were included, to avoid stars with 
highly uncertain proper motions. Furthermore, stars with proper motion
values larger than 15 mas/yr are eliminated as outliers. In this way, 
the sample of Sgr candidates consists of 156 objects. To determine their mean and dispersion, 
we have used the probability-plot method (Hamaker 1978)
using the inner $80\%$ of the proper-motion distribution. 
By doing so we aim to eliminate the minor contribution of distant 
halo stars to the estimate of the mean  
proper-motion of the candidates. The Besancon model (Robin et al. 2003) predicts 21 
halo stars in the CMD region and proper-motion range used to select
Sgr debris candidates. Therefore the chosen contamination fraction of $20\%$ seems to
slightly overestimate the Besancon model output. 
We find the mean relative proper-motion of Sgr debris candidates in SA 94 is 
$\mu_{\alpha}cos\delta = -0.18 \pm 0.32$ mas/yr and 
$\mu_{\delta} = -2.02 \pm 0.28$ mas/yr. 
By using fractions between $60\%$ and $90\%$, the estimate of the mean
did not change within its formal uncertainty. 
The mean absolute proper motion for candidate Sgr debris in SA 94
 is therefore $\mu_{\alpha}cos\delta = 0.14 \pm 0.36$ mas/yr and 
$\mu_{\delta} = -2.02 \pm 0.33$ mas/yr.
%In Galactic coordinates it is $\mu_{\l}cos~b = 1.52 \pm 0.35$ mas/yr and $\mu_{b} = -1.33 \pm 0.35$ mas/yr. 

For SA 93 we also present a preliminary value for Sgr stream candidates.
We plan to improve both determinations for SA 93 and 94 by obtaining
radial velocity data that will help establish membership to the stream.
For SA 93, the correction to absolute proper motion
is given by galaxies and QSOs together, rather than two 
independent measurements as
was done for SA 94. This is because 
there are only 2 QSOs measured on this field, one of which is a galaxy as well.
Proceeding in a similar manner as for SA 94, we obtain the correction to 
absolute proper motion
as given by the mean of 222 extragalactic objects:
$\mu_{\alpha}cos\delta = -1.96 \pm 0.40$ mas/yr and $\mu_{\delta} = 2.62 \pm 0.43$ mas/yr.
Since there are far fewer Sgr candidates in SA 93 than in SA 94, 
we have made use of all
stars within $r=14$ to 20 and $(g-r) = 0.0$ to 0.9 to determine the mean motion, 
since some subgiant and red clump stars may also be present besides the 
turnoff stars. In this case the halo contamination is much more important 
than in SA 94, and probability plots will not give an accurate result. 
We have thus simply selected the
candidates by defining a very conservative proper-motion cut: a circular region
of radius $\sim 2$ mas/yr that encompasses the proper-motion clump
seen in Figure 8. The mean is thus based on 63 stars, and it is:
$\mu_{\alpha}cos\delta = -2.54\pm 0.18$ mas/yr and 
$\mu_{\delta} = 0.30 \pm 0.24$ mas/yr.
The uncertainties are derived from the scatter given by the 63 Sgr 
candidates. We caution
that this uncertainty is formally low because it does not account for the 
uncertainty in choosing Sgr stream members. Finally, the absolute proper motion
for Sgr candidates in SA 93 is: 
$\mu_{\alpha}cos\delta = -0.58 \pm 0.47$ mas/yr and 
$\mu_{\delta} = -2.32 \pm 0.50$ mas/yr. We regard this number as 
preliminary, and with uncertainties possibly underestimated.
In both SA 94 and 93 determinations the membership to the stream is the 
major source of error. 

Majewski et al. (2006) have shown that by measuring the proper motion of
Sgr debris, especially along the trailing tail, one can 
determine the rotation velocity of the Local Standard of Rest (LSR), 
whereas Law et al. (2005) and
Johnston et al. (2005) show how the dynamics of the Sgr arms are also affected by the
flattening, $q$, of the Galactic potential. 
In Figures 9 and 10 we show proper-motion predictions by Law et al. (2005) for Sgr tidal debris
in the direction of SA 93 and SA 94 compared to the observed results found here
in these two fields.  The proper motions are shown as a function of 
the longitude along Sgr's orbit (Majewski et al. 2003); $\Lambda = 0^{\circ}$ 
corresponds to the main body of Sgr and $\Lambda$ increases in the trailing direction.
In Figure 9, the adopted velocity of the LSR is 220 km/s,
while the flattening of the halo $q$, has three values corresponding to 
prolate (top), spherical (middle) and oblate (bottom) halos.
In Figure 10, the adopted flattening of the halo is $q = 0.9$, and
the velocity of the LSR varies from 260 km/s to 180 km/s, as indicated
in each panel. The general agreement between our proper motion results and the model, 
which was constrained
solely from the distribution and radial velocities of M giants 
(Majewski et al. 2003, 2004) and a given Galactic potential 
(i.e., no proper-motion data), is remarkable.  This correspondence reinforces 
the assumption that we are 
measuring proper motions in Sgr's tidal stream. 
We note that the proper-motion 
range shown in Figs. 9 and 10 is only $\sim 3$ mas/yr, a typical value for the 
proper-motion uncertainty per well-measured star in catalogs such as Tycho2 (Hog et al. 2000), 
SPM 3 (Girard et al. 2004). 
To better illustrate this, we include here the mean absolute proper motion of 
red field stars (see also Fig. 8). In SA 94, we obtain $\mu_{\alpha}cos\delta = 2.80 \pm 0.57$ mas/yr and 
$\mu_{\delta} = -3.22 \pm 0.46$ mas/yr, and in SA 93,  $\mu_{\alpha}cos\delta = 3.99 \pm 0.88$ mas/yr
and $\mu_{\delta} = -0.94 \pm 0.80$ mas/yr.
Thus, along RA the mean 
proper-motions of red field stars in both areas lie outside the 
proper-motion range shown in Figs. 9 and 10.

\begin{figure}
\includegraphics[scale=0.90,clip=true]{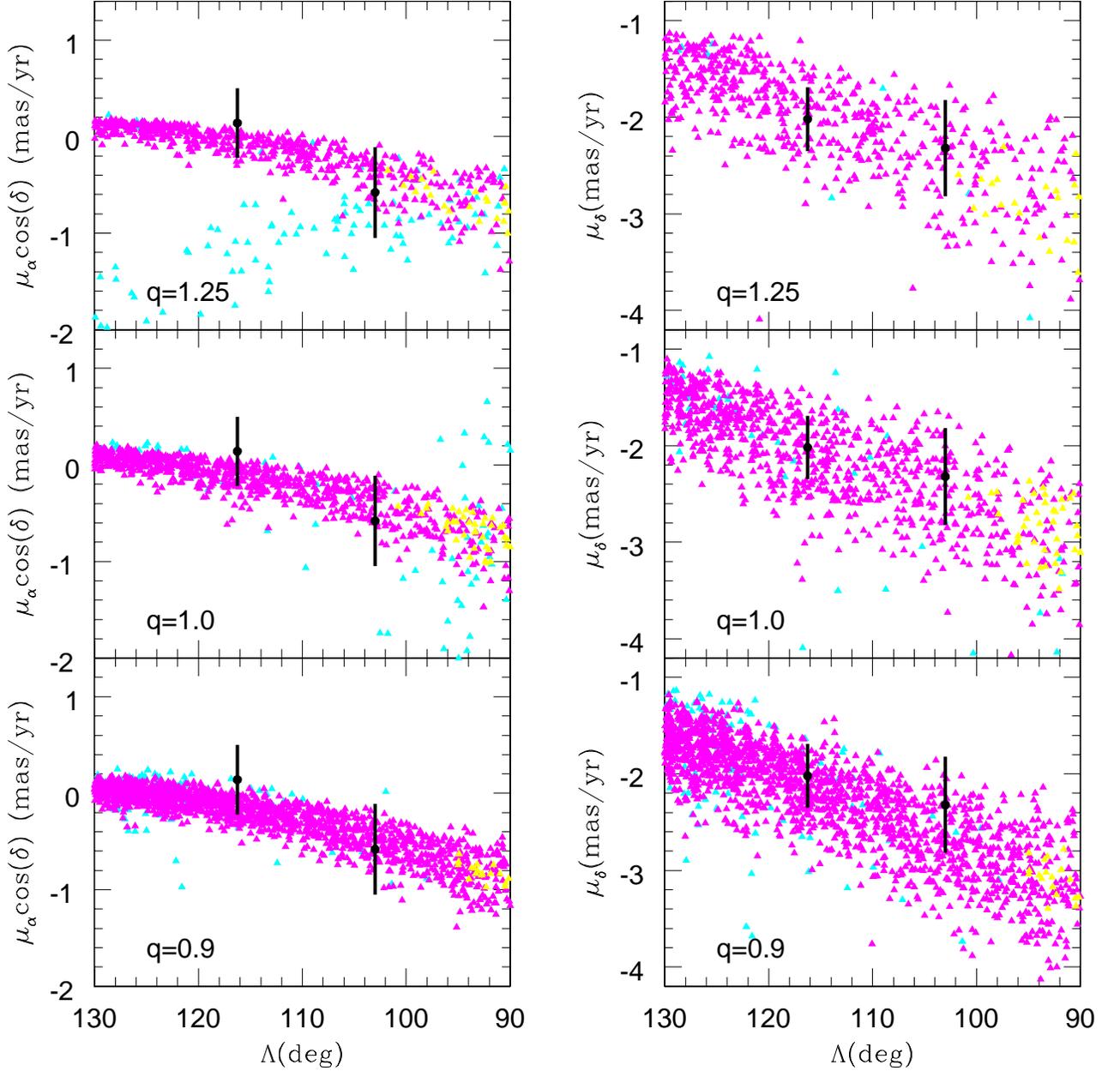}
\caption{Proper motions along Sgr's southern trailing tail as predicted
by the Law et al. (2005) models (color symbols). The dark symbols with
$1$-$\sigma$ error bars show our preliminary results in SA 93 
($\Lambda = 103^{\circ}$) and SA 94 ($\Lambda = 116^{\circ}$). The LSR velocity adopted
for this model is 220 km/s, while the flattening of the halo $q$ varies
as specified in each panel.  The colored dots represent N-body model particles stripped from Sgr
since its last apoGalacticon, i.e. present orbit, ({\it yellow symbols}), during the previous orbit ({\it magenta}),
and two orbits ago ({\it cyan}); this color scheme matches that used in Law et al. (2005)}
\end{figure}
\begin{figure}
\includegraphics[scale=0.90,clip=true]{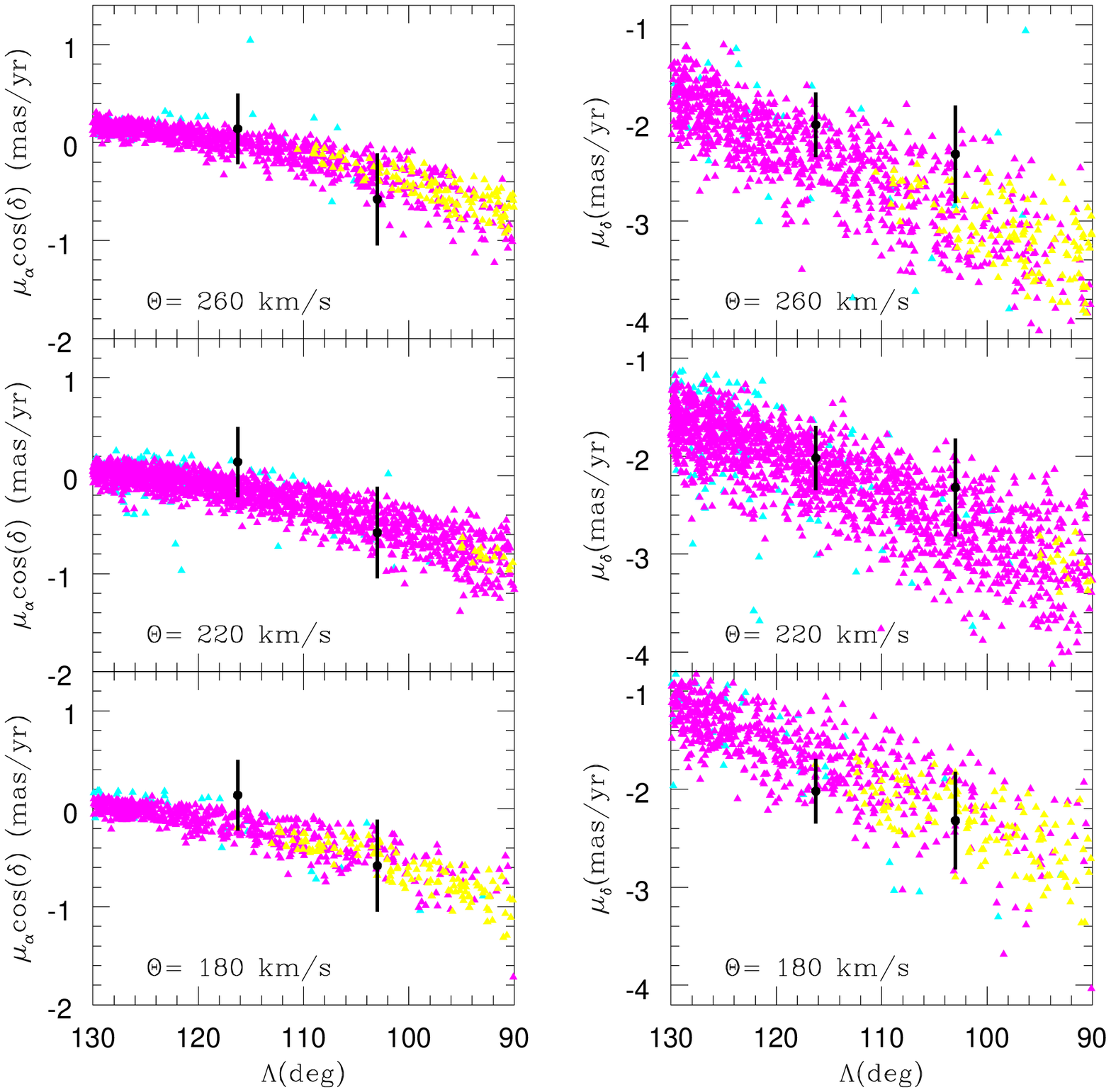}
\caption{Similar to Figure 9. The flattening of the halo adopted
for this model is 0.9, while the velocity of the LSR varies as specified 
in each panel.  Color representation is as in Fig.\ 9 and Law et al. (2005).}
\end{figure}

While our preliminary proper motions in only two SA fields do not yet lend discriminatory
power between the Galactic models shown, with improved proper motion samples and the inclusion
of additional fields at other $\Lambda$ (e.g., SA 116 and 117; see Fig.\ 1) we expect to be able
to more rigorously address this issue in the near future.

\subsection{The Monoceros Structure}

In Figure 11 we show similar plots to those in Figure 8 for the fields
SA 96 at $(l,b) = (198.3\arcdeg,-26.0\arcdeg)$,  
SA 100 at $(l,b) = (227.6\arcdeg,+26.7\arcdeg)$,  
and SA 101 at $(l,b) = (239.0\arcdeg,+39.9\arcdeg)$. 
These areas sample regions in the Monoceros structure
at the anticenter and across the Galactic plane. 
For SA 96, the data for the CMD are not in the available SDSS data releases.
We have obtained the data set from B. Yanny, and the photometry is
dereddened for this field. Reddening is however rather low for all three
of these fields: $E(B-V) = 0.07$ for SA 96,  $E(B-V) = 0.04$
for SA 100 and $E(B-V) = 0.04$ for SA 100.
Since the data for SA 96 are dereddened, we have shifted blueward
the color ranges
for selecting blue and red stars, by 0.1 mag compared to the other fields
(Fig. 11).
Newberg et al. (2002) designate the region containing SA 96 as
S200-24-19.8, for which their Fig. 15 shows a clear main sequence turnoff at $g = 19.8$. 
SA 100 is located only $3\arcdeg$ away from the eastern edge of
S223+20-19.4, which also displays a clear main sequence with the
turnoff at $g\sim 19.4$ (see Fig. 12 in Newberg et al. 2002). 
At $b \sim 40\arcdeg$, SA 101 is farther away from the Monoceros ring (see also Fig. 1).
The majority of the stars measured in our survey are however 
brighter than these turnoffs and the corresponding main sequence
stars that were studied by SDSS and  assigned to the Monoceros structure.

\begin{figure}
\includegraphics[scale=0.95,clip=true]{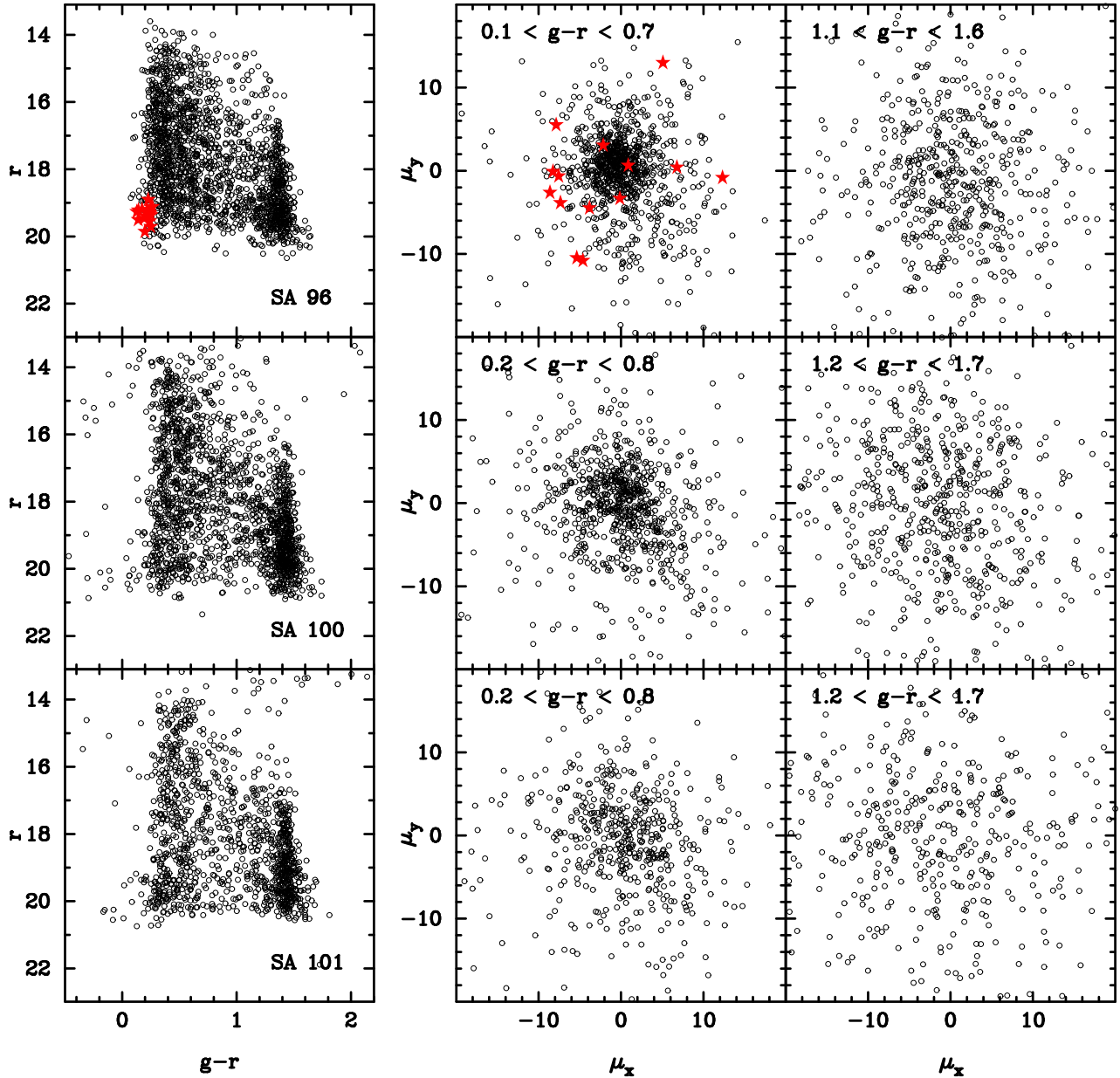}
\caption{Same as Fig. 8, only for SA 96, 100 and 101. The red symbols
in SA 96 are stars with radial velocities measured by Yanny et al. (2003, 2004)
in the turnoff of the Monoceros structure.}
\end{figure}

Yanny et al. (2003, 2004) measured radial velocities of candidate
turnoff stars in the Monoceros ring to demonstrate that their dispersion is
indicative of a kinematically cold stream (but cf. Crane et al. 2003). 
%
%SRM: I added the warning alert to Crane et al. because of the specific reference in the sentence
% to the Yanny et al. "cold dispersion".  Crane et al. not only points out that Yanny had significant
% zero-point errors in their RVs but also very largely inflated dispersions because they underestimated
% their RV errors.   I know this is not the point of bringing up the Yanny stars and would not object 
%to removing % the Crane reference as long as the part starting "to  demonstrate..." were removed.  Otherwise we are
% misleading the reader a bit that Yanny et al. actually did what you say correctly.
%
In Fig. 11, top row, the red
symbols are the stars in our SA 96 field that have radial 
velocities measured by
Yanny et al. (2003, 2004). These stars are at the faint limit 
of our survey, therefore their proper motions are quite uncertain.
Since the mean radial velocity of the candidate Monoceros stars  
overlaps with that of the thick disk (Yanny et al. 2003, 2004), 
it is also possible that some
of the stars are indeed thick disk stars, i.e., have
a larger proper-motion dispersion that that of a cold stream.
The tight clump seen in the proper-motion diagram of blue stars in SA 96
is comprised of stars with $r = 15$ to 19. Are these only thick disk/halo
stars? Comparing SA 96 with SA 100 which is located at the same
latitude as SA 96, only above the plane, and only $\sim 30\arcdeg$ 
away in longitude from SA 96, we sample 
very similar parts of the major Galactic components. However, there is a
clear difference in the number of blue stars in the two regions.
A careful inspection of Fig. 15 in Newberg et al. (2002)
which corresponds to SA 96 suggests that
multiple main sequences and turnoffs may be present, with the faintest 
one in SDSS being the most distinct.
To better illustrate the proper-motion clumpiness in SA 96,
we show a zoomed-in proper-motion diagram of the blue
stars (as defined in Fig. 11) for SA 96 (left panels) and SA 100
(right panels) in Figure 12. The middle and bottom panels show 
the proper motions as a function of magnitude.
The stellar excess as well as the proper-motion tightness in SA 96
are obvious when compared to SA 100.

\begin{figure}
\includegraphics[scale=0.90,clip=true]{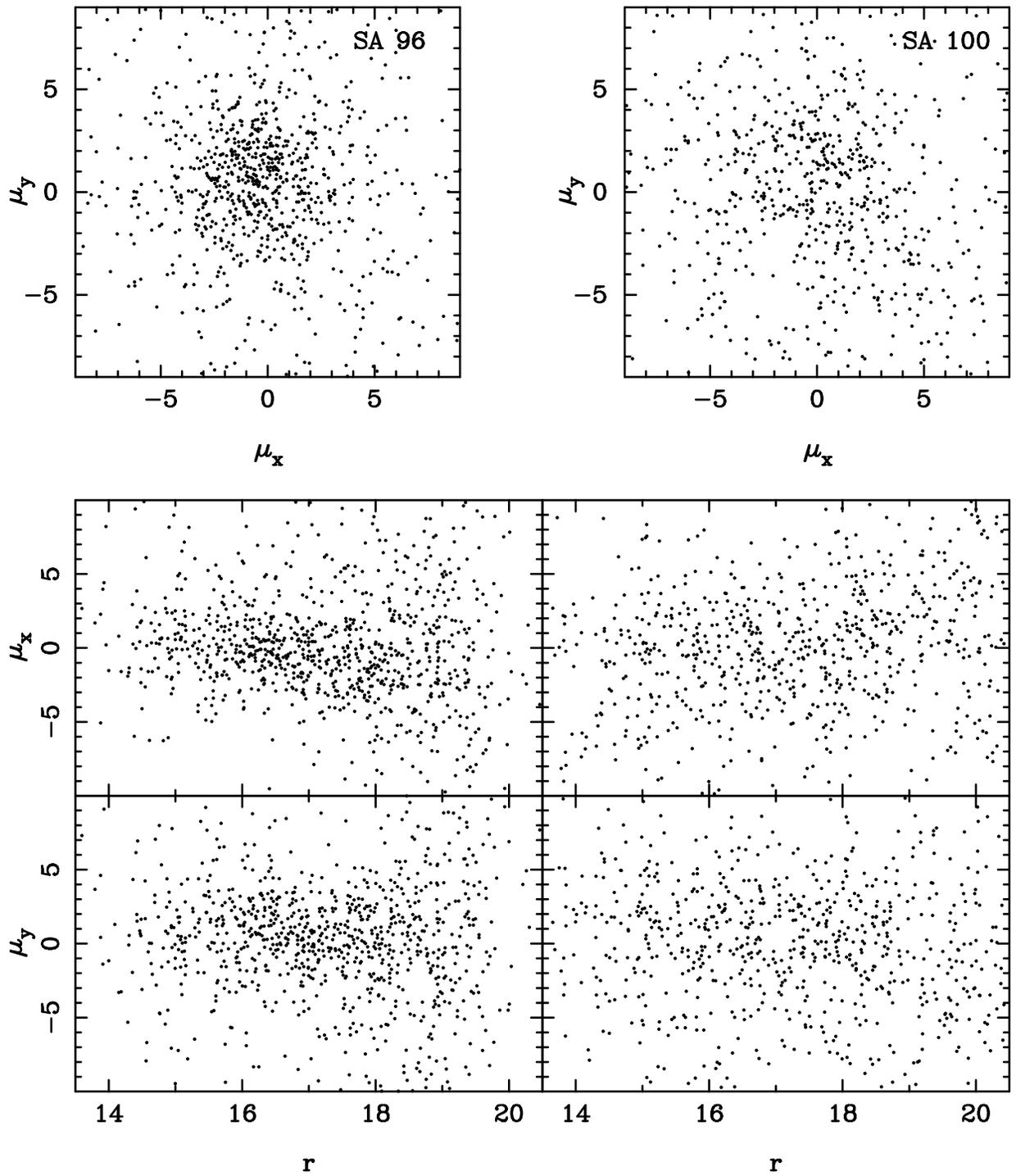}
\caption{Relative proper-motion distributions and proper motions as a
function of magnitude for blue stars in SA 96 (left panels)
and in SA 100 (right panels).}
\end{figure}

The stellar excess is also apparent when comparing the proper-motion
distributions in the three fields with those predicted by the 
Besancon Galactic model (Robin et al. 2003) which obviously contains
only the major Galactic components. 
In Figure 13 we show the absolute proper-motion distribution
along one coordinate (RA, for example)
 in all three fields as determined from the 
Besancon Galactic model (Robin et al. 2003) (top panels), and from our
data (bottom panels). The filter system of the Besancon models is the
CFHT MEGACAM system which is very close to the SDSS system. At any
rate, we are interested only in relative comparisons between fields,
rather than a direct comparison between data and model counts.
The model proper motions were convolved with a 1 mas/yr 
proper-motion uncertainty to approximately match the errors of the
observed proper motions.

\begin{figure}
\plotone{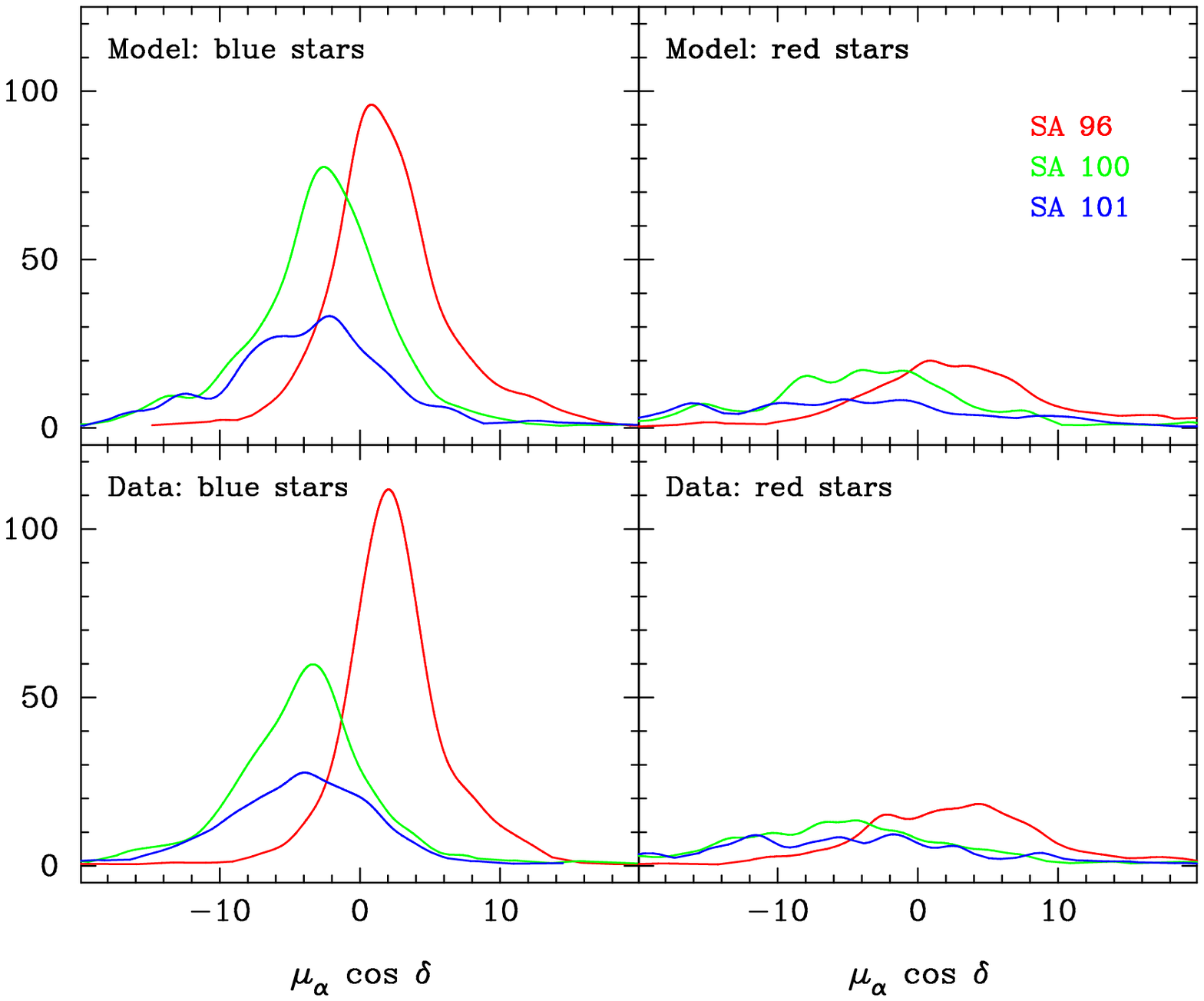}
\caption{Absolute proper-motion distribution along RA as given by
the Besancon Galactic model (top) and by our data (bottom). Blue and red
stars as selected in Fig. 11 are shown here in the left and right hand panels respectively. The magnitude range of the selected stars is $r=14$ to 19.}
%The red line is for SA 96, the green for SA 100, and the blue for SA 101.}
\end{figure}

The distributions were constructed for blue and red stars
as defined above (see also Fig. 11) and in the magnitude range $r = 14$ to 19.
SA 96 is represented with a red 
line, SA 100 with a green line and SA 101 with a blue line.
Rather than comparing directly the distributions given by the 
data with those given by the model for each field, we proceed to do a
relative comparison as follows. We compare pairs of fields, i.e. 
take ratios of the distributions as given separately 
by the data and the model for
SA 100 and SA 101, and for SA 96 and SA 100. These ratios are readily apparent
from the plot as given by the areas under each curve. Alternatively,
one can use the ratios of the peaks for pairs of fields.
By doing so, for the red stars it is apparent that the observations
are in good agreement with the model predictions.
For the blue stars, the ratios of the proper-motion distributions for
SA 100 and SA 101 as given by the model and the observations are in reasonably good agreement. 
However, this is not the case for the ratios for 
SA 96 and SA 100. Clearly the data show that SA 96 has an excess of stars when compared to the model predictions.
We also note here the good agreement between model and data for the
displacement in absolute proper motion between fields as the Galactic plane
is crossed.

One interpretation for the excess counts in SA 96 is that
in this region we see multiple, wrapped streams from the 
Monoceros structure, with the most distant one detected by SDSS at
$\sim 13$ kpc. Our preliminary absolute proper motion for candidate
Mon stars in SA 96 indicates a thick-disk-like motion in agreement
with the recently-modeled orbit of the Mon system
by Pe\~{n}arrubia et al. (2005).
Another possible interpretation
is that the excess stars in SA 96 are part of the Canis Major 
overdensity, under the assumption that this overdensity is
the core  of a disrupted dwarf galaxy.
The nature of this overdensity is strongly debated 
in the current literature. Other interpretations besides the dwarf galaxy
hypothesis are: the overdensity is the 
warp of the Galactic disk (e.g., Momany et al. 2006), a spiral
arm in the warped disk (e.g., Carraro et al. 2005), and the 
periphery of a system centered in the region of the Argo 
constellation (Rocha-Pinto et al.
2006). Currently, no interpretation of the reported CMa overdensity is clearly proved or 
widely accepted. If CMa proves to be any of the above interpretations
other than the dwarf galaxy one, then the excess in SA 96 is very unlikely to be
related to the CMa overdensity, primarily because of its location. 
We therefore explore the possible connection between SA 96 and the CMa 
overdensity interpreted as a dwarf galaxy.

In a recent wide-area study of the distribution of red clump stars
from 2MASS, Bellazini et al. (2006) map out the overdensity in
Canis Major, as well as the Galactic warp. From their study, CMa stands out
as a distinct overdensity in the outer regions of the Galactic disk.
More importantly, because it is so nearby, CMa covers a large portion of the 
sky, and the geometry is such that at lower longitudes, toward the anticenter,
CMa is closer to the Sun ($d_{\odot} \sim 6$ kpc) than its core,
 which is supposedly at $(l, b) = (240\arcdeg, -8\arcdeg), d_{\odot} \sim 8$ kpc 
(Bellazzini et al. 2006, Martinez-Delgado et al. 2005). 
Two figures in Bellazzini et al. (2006) support the notion that SA 96 may be 
on the outskirts of the reported CMa overdensity: (a) their Figure 9 (top panel), which shows the 
excess of surface density of CMa and indicates that CMa is 
extended away from the Galactic plane at longitudes closer to the Galactic
Anticenter; and (b) Figure 13 (bottom panel) of Bellazini et al., which shows
a Sgr-type galaxy placed at the distance and location of CMa. 
Indeed its putative extension is impressive, and SA 96 would lie at the 
tidal radius of such a system.
Another indication that the system may be elongated in the direction of
Galactic rotation and away from the plane is the motion of CMa as
measured by Dinescu et al. (2005b): the $\Theta$ velocity component is
$188 \pm 15$ km/s, while the $W$ component is $-49 \pm 15$ km/s.
Radial velocities of our proper motion stars in the field of SA 96 should help clarify this 
issue and hopefully lend further clues to the nature of Mon and CMa structures.

We proceed now to estimate the proper-motion uncertainty by assuming that
the blue stars in the proper-motion clump seen in SA 96 belong to
a kinematically cold system, and therefore their proper-motion dispersion 
reflects measurement errors only. 
In the proper-motion diagram of SA 96,
in Fig. 12 there appear to be two overlapping proper-motion clumps with different dispersion: 
one less tightly clumped, centered at 
$(\mu_x, \mu_y) \sim (0,0)$ mas/yr and a ``radius'' of $\sim 4$ mas/yr, 
and the other more tightly clumped centered at
$(\mu_x, \mu_y) \sim (-1,+2)$ mas/yr and a ``radius'' of $\sim 2$ mas/yr.
Selecting stars within the radius of the less tight clump, and within $r= 14$ to 18,
we obtain a proper-motion scatter of 1.6 mas/yr. Similarly, for the tighter clump
we obtain a proper-motion scatter of $\sim 1$ mas/yr. These numbers are only
approximate estimates to illustrate the precision of the proper motions in this work.
A rigourous kinematical understanding and therefore a better separation 
of these two clumps is beyond the scope of this paper. This will be addressed in future
work where radial velocities will be considered also.
One more cautionary note should be made:
the population of the
less tight proper-motion clump seems to appear as well in SA 100 (Fig. 12), and may 
therefore be representative of distant halo/thick disk stars, i.e., a population with a
non-negligeable intrinsic proper-motion dispersion.

%Other fields that we have reduced and that
%indicate structure in the proper-motion space are SA 71
%and SA 72. Excess star counts and proper-motions in SA 71 were 
%previously presented by our group (Dinescu et al. 2002). At that time
%the data in SA 71 were interpreted as debris from Sgr; however preliminary
%radial velocities in SA 71 indicate that there are only a few stars that
%belong to Sgr, and these are toward the faint end ($V > 19$) of the survey. 
%According to the proper motions and radial velocities, the majority
%of the excess stars found in SA 71 within $V = 17$ to 18, more likely belong 
%to Mon. This will be readdressed in a future paper with more complete data.

\section{Summary}

We describe our ground-based, photographic 
absolute proper-motion survey along three near-equatorial declination 
zones. Proper motions are derived in $40\arcmin\times40\arcmin$ fields
from a collection of 2.5-m Du Pont, 4-m Mayall, 60-inch Mt. Wilson and
POSS-I plates. The time baseline varies between 40 and 85 years.
We have demonstrated that we obtain proper-motion uncertainties 
between $\sim 1$ and 3 mas/yr per star down to a magnitude of 19, 
and for a few fields down to 21. The typical uncertainty in the 
correction to absolute proper motions as given by galaxies and QSOs is
between 0.2 and 0.8 mas/yr.
The described proper-motion survey is complemented by our own ongoing
radial-velocity and photometric follow-up programs 
as well as by current surveys such as SDSS, QUEST, etc.
The characteristics of this proper-motion survey make it suitable to address
many topics related to both the main Galactic components, and the
tidal features seen in the halo and associated with streams from disrupted
satellites.
In this contribution, we present preliminary results in the southern 
trailing tidal tail of Sgr and in the Monoceros ring region.

We thank Stephen Levine from USNO-Flagstaff for making available the 
POSS-I USNO scans.
SRM is grateful to the Carnegie Observatories and its former director Augustus Oemler
for a Carnegie Visiting Associateship that made possible the collection of the Du Pont plates 
used in this survey.
The financial support from NSF grants AST-0406884
and AST-0407207 for this research is acknowledged.

This publication makes uses of SDSS data products.
Funding for the SDSS and SDSS-II has been provided by the Alfred P. Sloan 
Foundation, the Participating Institutions, the National Science Foundation, 
the U.S. Department of Energy, the National Aeronautics and Space 
Administration, the Japanese Monbukagakusho, the Max Planck Society, 
and the Higher Education Funding Council for England. The SDSS Web Site 
is http://www.sdss.org/.
%The SDSS is managed by the Astrophysical Research Consortium for the 
%Participating Institutions. The Participating Institutions are the American 
%Museum of Natural History, Astrophysical Institute Potsdam, 
%University of Basel, Cambridge University, Case Western Reserve University, 
%University of Chicago, Drexel University, Fermilab, the Institute for 
%Advanced Study, the Japan Participation Group, Johns Hopkins University, 
%the Joint Institute for Nuclear Astrophysics, the Kavli Institute for 
%Particle Astrophysics and Cosmology, the Korean Scientist Group, the 
%Chinese Academy of Sciences (LAMOST), Los Alamos National Laboratory, 
%the Max-Planck-Institute for Astronomy (MPIA), the Max-Planck-Institute 
%for Astrophysics (MPA), New Mexico State University, Ohio State University, 
%University of Pittsburgh, University of Portsmouth, Princeton University, 
%the United States Naval Observatory, and the University of Washington.

\end{document}